\documentclass[10pt, conference, letterpaper]{IEEEtran}

\usepackage[english]{babel}
\usepackage{graphicx}
\usepackage{nicefrac}
\usepackage{cite}
\usepackage{blindtext}
\usepackage{subcaption} 
\usepackage{algorithm}
\usepackage{algorithmicx} 
\usepackage[noend]{algpseudocode}
\usepackage{soul}
\usepackage{array}
\usepackage{multirow}
\usepackage{enumitem}
\usepackage{tablefootnote}
\usepackage{amsthm,amsmath,amssymb,amsfonts}
\usepackage{url}
\usepackage{eso-pic}

\newtheorem{remark}{Remark}[section]

\begin{document}
\AddToShipoutPictureFG*{\AtPageLowerLeft{\raisebox{0.28in}[0pt][0pt]{\makebox[\paperwidth][c]{\parbox{0.94\paperwidth}{\centering\scriptsize This work has been submitted to the IEEE for possible publication. Copyright may be transferred without notice, after which this version may no longer be accessible.}}}}}
\title{RetryGuard: Preventing Self-Inflicted and Attack-Driven Retry Storms in Cloud Applications}

\author{
\IEEEauthorblockN{
Jhonatan Tavori\IEEEauthorrefmark{1} \;\;
Anat Bremler-Barr\IEEEauthorrefmark{4} \;\;
Hanoch Levy\IEEEauthorrefmark{4} \;\;
Ofek Lavi\IEEEauthorrefmark{4}
}
\IEEEauthorblockA{\IEEEauthorrefmark{1}Columbia University \quad \quad \IEEEauthorrefmark{4}Tel Aviv University}
}

\maketitle

\begin{abstract}
Modern cloud applications are built from independent microservices, offering scalability and usage-based billing. However, their reliance on independently-operating auto-scalers introduces coordination challenges. 
Default retry patterns can trigger ``retry storms'' during service miscoordination or adversarial overload, amplifying load, latency, and resource billing. These storms can cause either self-inflicted Denial-of-Wallet (DoW) or amplify the impact of DDoS attacks.

To overcome these problems, we introduce \textit{RetryGuard}, a distributed framework for productive control of retry patterns across interdependent microservices. By managing retry policy on a per-service basis and making parallel decisions, \textit{RetryGuard} prevents retry storms, curbs resource contention, and mitigates escalating operational costs. \textit{RetryGuard} makes its decisions based on an analytical model that captures the relationships among retries, throughput (rejections), delays, and costs. 
Simulations show that it outperforms established mechanisms, including exponential backoff, jitter, and retry budgets.

Experimental results show that \textit{RetryGuard} significantly reduces resource usage and costs compared to existing AWS policies, achieving more than \(90\%\) improvement in latency, and \(98\%\) reduction in storm size. We further demonstrate its effectiveness in a multi-layer Kubernetes deployment with the Istio service-mesh, where \textit{RetryGuard} reduces the peak number of replicas by \(3 \times\) and cumulative memory usage by \(55\%\).

\end{abstract}

\section{Introduction}
Retries are among the oldest fault-tolerance techniques in networking, dating back to ARPANET \cite{ lee1988optimal}. In today’s cloud environments, application-level retries, typically layered atop HTTP transports such as REST and gRPC, are enabled by default in most platforms \cite{microsoft_retry_storm,googleselfinf}.
Yet, as demonstrated in this work, common cloud retry policies can trigger costly self-inflicted retry storms in microservice-based applications.

In recent years, cloud applications have shifted from monoliths to microservices-based architectures, decomposing functionality into smaller independent services with clear interfaces \cite{gan2018architectural,kratzke2018brief}, to improve scalability, manageability, and deployment flexibility
\cite{rahman2019predicting,xing2021charon,grewal2023expressive}.
Thanks to \textit{auto-scaling} mechanisms, microservices can be deployed across diverse hardware configurations and are billed based on usage (e.g., per operation or per runtime), aligning with the \textit{serverless} approach.

At face value, it appears as if microservices applications, supported by auto-scaling capabilities, would free operators from concerns about manual resource allocation \cite{mampage2022holistic}.
Nonetheless, this flexibility introduces  significant coordination challenges, posed by the interactions between independently developed and deployed services whose operation is inter-dependent. 
Consequently, the services may experience periods of {\it miscoordination}\footnote{We use \textit{miscoordination} to denote a prolonged mismatch between the request rate of one tier and the serving capacity of another. This can arise from delayed scaling (e.g., an upstream service scales faster than its downstream dependency) or misconfigurations (e.g., manual scaling or concurrency limits).}, lasting several minutes, during which they may be subject to blocking (rejections) and delays. 
During periods of miscoordination, retry operations, which are the focus of this work, may become counterproductive. 
\\
\\
\textbf{Retry Operation under Miscoordination.}
Retries between services in tandem have been designed to perform the important task of overcoming {\it instantaneous} service blocking (at the downstream service) due to statistical variations of the request stream or random service errors. Their 
effectiveness generally relies on the services being well coordinated,
namely the downstream service  operates under  \textit{stability conditions} with load
$\rho < 1 $, where $\rho$ is the total arrival rate, including retries, divided by the downstream service rate.
Our key observation, however,  is that when the services enter a period of {\it prolonged miscoordination}, 
during which the downstream  service experiences  load  
$\rho >1 $, their  behavior changes drastically, and retries become counterproductive, exacerbating inefficiencies and inflating operational costs.

Such miscoordination is common in modern microservices-based 
cloud applications and may arise from: (1) \textit{Independent auto-scalers}, causing delays in matching capacities between service tiers; (2) \textit{Misconfiguration} errors, 
leading to
persistent miscoordination 
\cite{zheng2007automatic}; and (3) \textit{DDoS attacks}, which impose sustained overload through high volumes of malicious traffic.

\begin{figure}[h]
    \centering
    \includegraphics[width=0.75\linewidth]{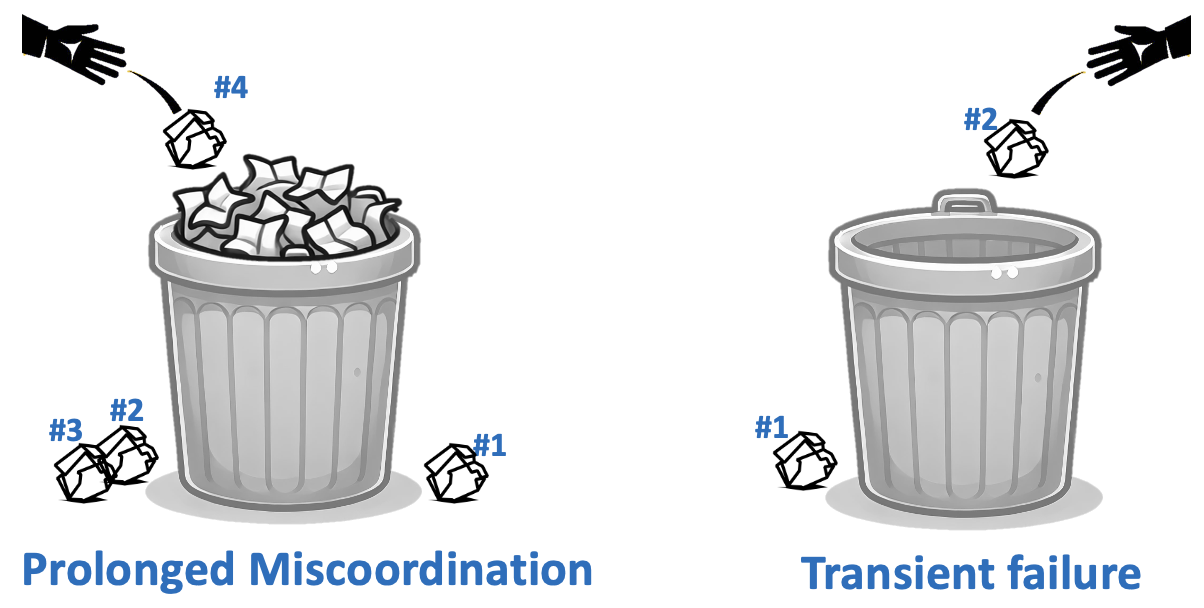}
    \caption{Prolonged versus instantaneous blocking.}      \label{fig:figure1}
\end{figure}

Fig.~\ref{fig:figure1} illustrates the target service as a trash basket. Under stable load, retries are likely to succeed; under prolonged miscoordination, unavailable capacity causes repeated failures, further increasing load, delay, and rejection.
\\

\begin{figure}[h]
    \centering
    \includegraphics[width=\linewidth]{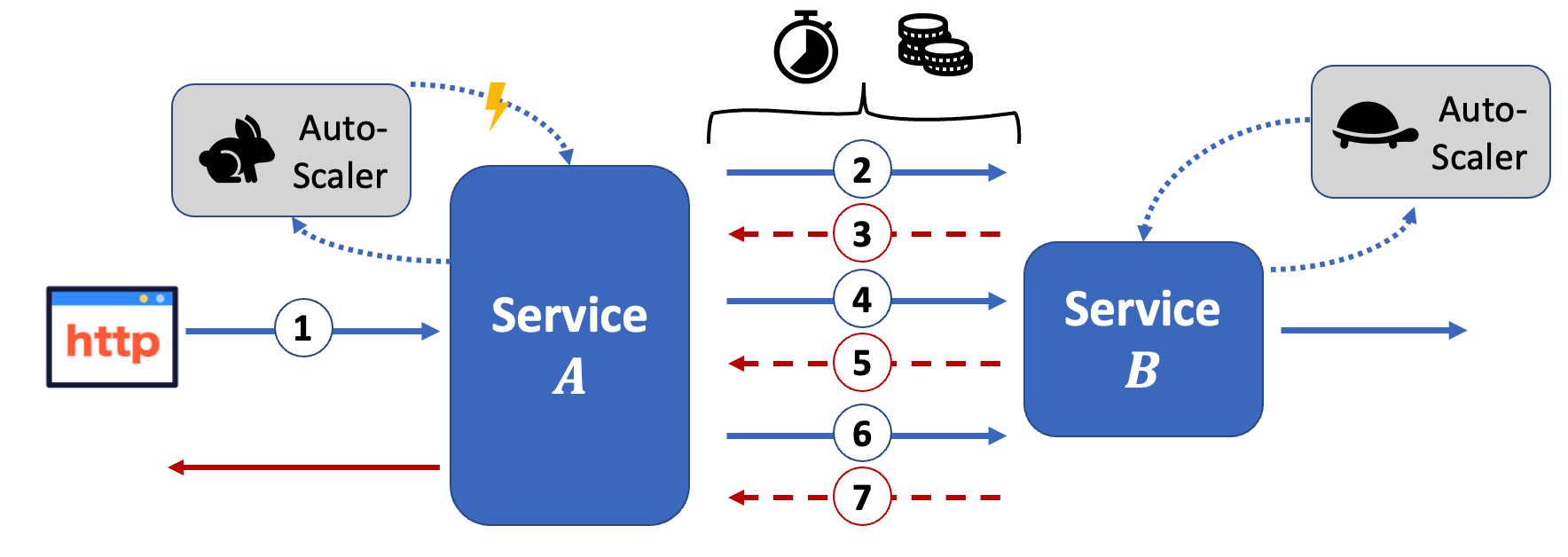}
    \caption{Self-Inflicted Retry Storm: Miscoordinated tandem services.}
        \label{fig:figure2}
\end{figure}

\noindent
\textbf{Retry Storms: Self-Inflicted or Attack-Driven DoW.}
Excessive retries can escalate into costly and disruptive “retry storms”, a form of self-inflicted Denial-of-Service (DoS) scenario\footnote{Self-inflicted DoS occurs when a service generates more traffic than it can handle, resulting in throttling, delays, or failures.} 
\cite{microsoft_retry_storm, linkerd_retries_go_wrong}.
Figure \ref{fig:figure2} illustrates a retry storm.  
If Service \(A\)'s auto-scaler  reacts faster than Service \(B\)'s, a surge in HTTP requests causes \(A\) to scale up while \(B\) lags behind, creating a \textit{bottleneck}.  
As a result, blocked requests at \(B\) may accumulate at \(A\), forming a \textit{wave} awaiting completion.
Service \(A\), assuming the issue is temporary, retries the blocked requests multiple times, causing a snowball effect on the already struggling Service \(B\).
Each failed request consumes resources across multiple retry attempts before ultimately failing.
Furthermore, when \(B\) eventually scales, it may overestimate demand based on the accumulated wave.  
Thus, the system experiences a self-inflicted Denial-of-Wallet (DoW) scenario: Retry-amplified requests load in the upstream services, as well as over-scaling at the downstream service. Both phenomena, which may last for a long duration, inflict significant  costs on the application. 

The same amplification can be triggered externally: A short DDoS burst can overload the downstream service, with retries sustaining the storm after the attack ends and turning it into a prolonged DoW condition.
\\

\noindent
\textbf{\textit{RetryGuard}: Optimizing Retries Control.}
Retry-amplified traffic highlights the need to adapt retry strategies to situations of miscoordination  between services and to detect when retries become counterproductive. This adaptation is essential not only for maintaining stability and reducing costs, but also for improving overall performance.
We present {\bf \textit{RetryGuard}}, a distributed mechanism designed to mitigate self-inflicted retry storms based on this principle. 
\textit{RetryGuard} dynamically detects miscoordination, based on various parameters (including retry volume, rejections, or delays), and enables or disables retries at the service level accordingly. Our approach provides a generalizable framework applicable across diverse systems.

Unlike common retry strategies, including exponential backoff \cite{kwak2005performance}, jitter \cite{aws_jitter}, retry budgets \cite{linkerd_traffic_managment}, and leaky buckets \cite{butto1991effectiveness}, \textit{RetryGuard} enforces fundamentally distinct operational rules 
for transient failures and prolonged miscoordination.
By detecting sustained mismatches between service tiers, it suppresses retries only when they become counterproductive, offering a fundamentally different approach from legacy mechanisms. Their behavior and limitations are examined in Sections \ref{sec:retrydesign} and \ref{sec:evaluation}. Compared with circuit breakers,  \textit{RetryGuard} targets the retry storm itself by suppressing only retries while allowing new requests to use the available capacity.

\textit{RetryGuard} is supported by traffic analysis of a tandem system (Sec. \ref{sec:analysis}), which characterizes traffic properties and yields sharp differentiation between the situations of instantaneous failures and prolonged miscoordination.
Experiments (Sec. \ref{sec:evaluation}) demonstrate its advantages over existing retry policies.

\begin{table}[t]
\centering
\renewcommand{\arraystretch}{1.25}
\begin{tabular}{|>{\centering\arraybackslash}m{0.2cm}||p{1.45cm}||p{1.2cm}|p{1.7cm}|p{2.05cm}|}
\hline
\multirow{6}{*}{\centering\rotatebox{90}{\textbf{AWS}}} 
& {Retry Policy}  & Retries/Req & Rejection Rate & Resources Billing
\\ \cline{2-5}
& Legacy       & $2.09$       & $19.8\%$           &  $10.29\times  $          \\ \cline{2-5}
& Standard           & $1.08$         & $19.4\%$     &  $3.29\times$                           \\ \cline{2-5}
& Adaptive                        & $0.52 $     & $19.6 \%$            &  $1.67\times$                         \\ \cline{2-5}
& \textit{\textbf{RetryGuard}} & \textbf{0.05}   & \textbf{18.8\%}  & \textbf{1.00}$\times$    \\ \hline
\end{tabular} 

\vspace{0.2cm} 

\begin{tabular}{|>{\centering\arraybackslash}m{0.2cm}||p{1.45cm}||p{1.2cm}|p{1.7cm}|p{2.05cm}|}
\hline
\multirow{3}{*}{\centering\rotatebox{90}{\textbf{Istio}}} 
& {Retry Policy}  & Retries/Req & Rejection Rate & Peak Replicas
 \\ \cline{2-5}
& Default &  0.31 
& $5 \%$  & $616$ \\
\cline{2-5}
& \textbf{\textit{RetryGuard}} &  \textbf{0.01 }
&  \textbf{4\%} & \textbf{209} \\
\hline
\end{tabular}

\vspace{0.1cm} 

\caption{Eliminating Superfluous Costs: Comparison of \textit{RetryGuard} with AWS and Istio retry policies under 150--200\% load during scaling. \textit{RetryGuard} substantially reduces retry amplification, resource consumption, and associated costs while maintaining or improving the rejection rate.}
\label{tab:retryguard_comparison_combined}

\end{table}

In Table \ref{tab:retryguard_comparison_combined}, we present a 
summary of \textit{RetryGuard}’s performance in comparison to AWS and Istio retry policies, with a full evaluation detailed in Sec. \ref{sec:evaluation}. It
summarizes \textit{RetryGuard}'s performance\footnote{Note that, naturally, some request failures are inevitable due to the capacity limits of bottleneck services (i.e., the pigeonhole principle) regardless of the retry strategy. However, \textit{RetryGuard} minimizes additional failures.} and resource-cost improvements.

\subsection{Main Contributions}
\noindent
We make the following key observations:
\begin{itemize}
\item Retries present challenges that are intrinsically linked to modern cloud applications.
\item Under prolonged miscoordination and overload,  retries turn out to be counterproductive. Existing retry solutions are ineffective in miscoordination scenarios.
\item Economic considerations are critical in this context, while prior efforts have predominantly focused on performance.
\end{itemize}
\noindent
The main contributions of this paper are as follows:
\begin{itemize}
\item Demonstrating self-inflicted DoW using an AWS Serverless application and an Istio-based Kubernetes application, showing its impact on latency, cost, CPU usage, and the number of provisioned replicas (Sec.~\ref{sec:3}).
\item Proposing and implementing \textbf{\textit{RetryGuard}}, a distributed mechanism based on a productive-retry controller (see Sec.~\ref{sec:retrydesign} for design principles), and comparing it with exponential backoff, jitter, and retry budgets.
\item Providing a mathematical model and cost analysis for the core of the problem (Sec.~\ref{sec:analysis}), deriving the activation threshold for the productive-retry controller, and supporting its robustness and feasibility.
\item Deploying and experimentally evaluating \textit{RetryGuard} against AWS's  retry APIs and Istio's retry mechanisms in a Kubernetes service-mesh deployment (Sec.~\ref{sec:evaluation}).
\item Examining how retry mechanisms can be exploited by malicious users through DDoS attacks to trigger over-provisioning and prolonged request lifetimes, and discussing \textit{RetryGuard} as a mitigation strategy (Sec.~\ref{sec:discussion}).
\end{itemize}

Additionally, we provide an overview of key cloud challenges, including coordination difficulties and the limitations of existing solutions (Sec.~\ref{sec:background}).
In Sec. \ref{sec:discussion} we discuss the economic \textit{interplay} between the cloud provider and the application owners (the cloud customers); concluding remarks are given in Sec. \ref{sec:conclusions}.

\section{Background and Related Work}\label{sec:background}

We discuss
the limitations of existing mechanisms
in handling retries under miscoordination and related prior work.

\subsection{Background}

\noindent
\textbf{Retry Mechanisms Are Integral to Cloud Applications.}
Retry mechanisms handle failures by transparently retrying failed requests for service \cite{azure_retries_pattern}, and have been used for decades as a resiliency mechanism in fault tolerant computer systems \cite{lee1988optimal} and networks \cite{azure_retries, linkerd_traffic_managment, google_exponential_backoff}.
Microservices frequently interact over HTTP, HTTPS \cite{yu2019survey} or RPC \cite{nikita2021} protocols, and the retries are embedded in the calling functions (provided by the SDKs). Many providers incorporate retry mechanisms into their libraries, offering automated retry implementations, and recommend traditional techniques like exponential backoff and jitter (e.g., AWS \cite{aws_jitter} and Microsoft Azure \cite{azure_retries}).
This approach is also common in service mesh platforms \cite{istio_bookinfo,linkerd_retries_go_wrong}. A key feature of these platforms is their ability to automatically handle retries for failures \cite{saleh2022empirical, ashok2021leveraging}. According to a recent survey by the CNCF, 56\% of respondents identified reliability features, such as  retries, as a key driver for adopting service mesh technology \cite{ccnfreport}.
\\

\noindent
\textbf{Collaborating Microservices Face Coordination Challenges}. 
When multiple microservices collaborate to handle high-volume traffic, ensuring smooth coordination becomes a critical challenge, particularly if each service follows its own scaling policy. 
Auto-scaling, a key feature of cloud computing, automatically adjusts resources to accommodate changing workloads \cite{vaquero2011dynamically} and offers significant flexibility.
However, since  different services may scale in different ways \cite{liu2015prorenata, chen2018survey}, this flexibility may also introduce miscoordination. For example, \textit{databases} usually scale vertically (by provisioning larger instances) or through sharding, \textit{virtual machines} rely on resizing or load balancing, and \textit{Functions-as-a-Service} (FaaS) scale almost instantly.
Miscoordination between the scaling of the services may lead to resource waste and increased budget consumption. Similar dependencies also arise in service meshes \cite{michaelis2024l3}.
\cite{bremlerexploiting} highlighted budget waste caused by miscoordination, showing that customers often pay for doomed-to-fail requests. However, they \textit{did not} address retries.
\\

\noindent
\textbf{Cost Tracking in the Cloud is Challenging.}
It is important to note that understanding cloud spending remains challenging 
due to complex billing and applications spanning hundreds or thousands of microservices~\cite{walterbusch2013evaluating}. 
Organizations frequently struggle to identify high-cost services, resulting in potential overspending\footnote{Obtaining precise, time-specific cost data remains challenging as metrics tools often aggregate usage into coarse intervals \cite{amazoncloudwatch} and some report only every few minutes, limiting visibility into sudden fluctuations.}.
In the scenarios highlighted in this paper, 
the lack of granular accurate tracking makes it much harder to monitor expenses and detect inefficiencies or costs caused by miscoordination and retry storms, often leaving operators unaware of spikes until after significant charges have 
accumulated.

\subsection{Related Work}
Optimization of cloud microservice architectures has been the focus of extensive recent work, with techniques such as circuit breakers, timeouts, and load controls.  
Prior work has addressed overload \cite{iyer2001overload} through per-caller backoff \cite{jacobson1988congestion}, server-side admission control and token-based schemes \cite{Ren18DAGOR,cho2020overload,xing2025rajomon}, reinforcement learning \cite{park2024topfull}, and credit-based admission \cite{cho2023protego}.  Other works focused on server-side issues like CPU contention \cite{cho2020overload,welsh2002overload}.
These approaches primarily regulate offered load and do not specifically address retry storms caused by service miscoordination. \textit{RetryGuard} instead removes the self-inflicted, retry-amplified portion of the load while preserving first attempts, leaving the underlying capacity shortfall to autoscaling or existing overload controllers. Thus, it is complementary to mechanisms such as~\cite{cho2020overload,cho2023protego,huang2022metastable}.

Retry mechanisms are widely supported by cloud providers and service-mesh platforms, which offer guidance for avoiding common pitfalls, including Microsoft~\cite{microsoft_retry_storm}, Google Cloud~\cite{googleselfinf}, and Linkerd~\cite{linkerd_retries_go_wrong}.
These recommendations typically control retry timing or rate.
Exponential backoff~\cite{kwak2005performance} and jitter~\cite{aws_jitter} delay and spread retries under the assumption that failures are transient, while Linkerd’s retry budgets \cite{linkerd_traffic_managment} limit the retry rate. However, as discussed in Sec.~\ref{sec:retrydesign}, these mechanisms are ineffective under prolonged service-level miscoordination, where failures persist and the retries themselves become harmful. \textit{RetryGuard} adds regime detection, treating sustained miscoordination as a distinct operating mode in which retries are suppressed while first attempts are preserved.

{Despite} the recent academic interest in microservices-based architectures, limited research exists on how retry patterns impact performance in cloud-based applications. 
{Recently}, \cite{saleh2022empirical} conducted experiments on a Kubernetes testbed using the Istio service mesh, demonstrating that retry configurations significantly influence application throughput and latency. \cite{mendonca2020model,costofretrying} showed similar findings using a probabilistic model checker.
However, to the best of our knowledge, no systematic study has been conducted to study the behavior and economic impact of retries in the context of microservices miscoordination.

\section{Case Studies: Self-Inflicted DoW}\label{sec:3}
To demonstrate self-inflicted DoW resulting from retry storms, and its severity, we present two case studies: an AWS web application and a Kubernetes application with Istio service mesh.
Both cases will demonstrate how, during miscoordination and overload conditions, retries can significantly \textbf{increase} resource billing costs for the application owner \textit{while} potentially severely \textbf{degrading} user performance.

\subsection{AWS Lambda and DynamoDB}\label{subsec:3-aws}
We choose a popular AWS architecture, consisting of 
\textit{AWS Lambda}, a serverless compute service (\textit{FaaS})  
\cite{aws_lambda_docs}, and
\textit{AWS DynamoDB}, a popular fully-managed 
database 
\cite{aws_dynamodb_docs}.

\subsubsection{Setup}
\begin{figure}[h]
    \centering
	\includegraphics[width=0.85\linewidth]{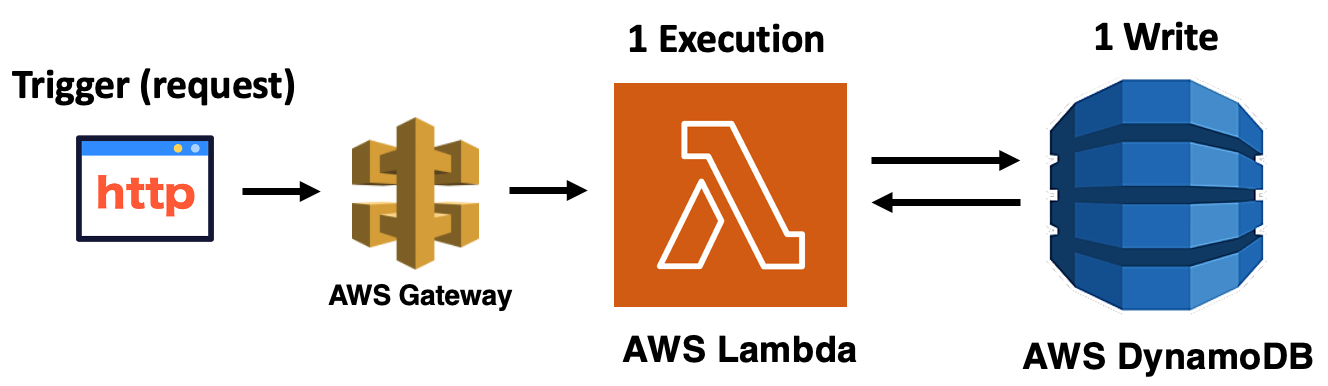}
	\caption{\textit{AWS Lambda} and \textit{DynamoDB} in tandem.}
	\label{fig:lambdadb}
    \vspace{0.2em}
\end{figure}

Each HTTP request triggers a single Lambda execution that performs a write operation to DynamoDB. The tested architecture is depicted in Fig. \ref{fig:lambdadb}.
Lambda and DynamoDB scale differently: 
Lambda scales nearly \textit{instantaneously} by provisioning a separate instance for each request \cite{aws_lambda_docs}.
In contrast, DynamoDB (in provisioned mode) attempts to adapt to the measured arrival rate. This process scaling involves a \textit{delay} due to the decision-making process and resource adjustment \cite{aws_dynamodb_docs}. 
Throttled DynamoDB requests 
returned an error, prompting Lambda to decide on a retry. The Lambda function used 
AWS default retry configuration.

\subsubsection{Self-Inflicted DoW: Over-Scaling and Extensive Delays}

\begin{figure}[t]
    \centering
    \begin{subfigure}[c]{\linewidth}
        \centering
        \includegraphics[width=0.87\linewidth]{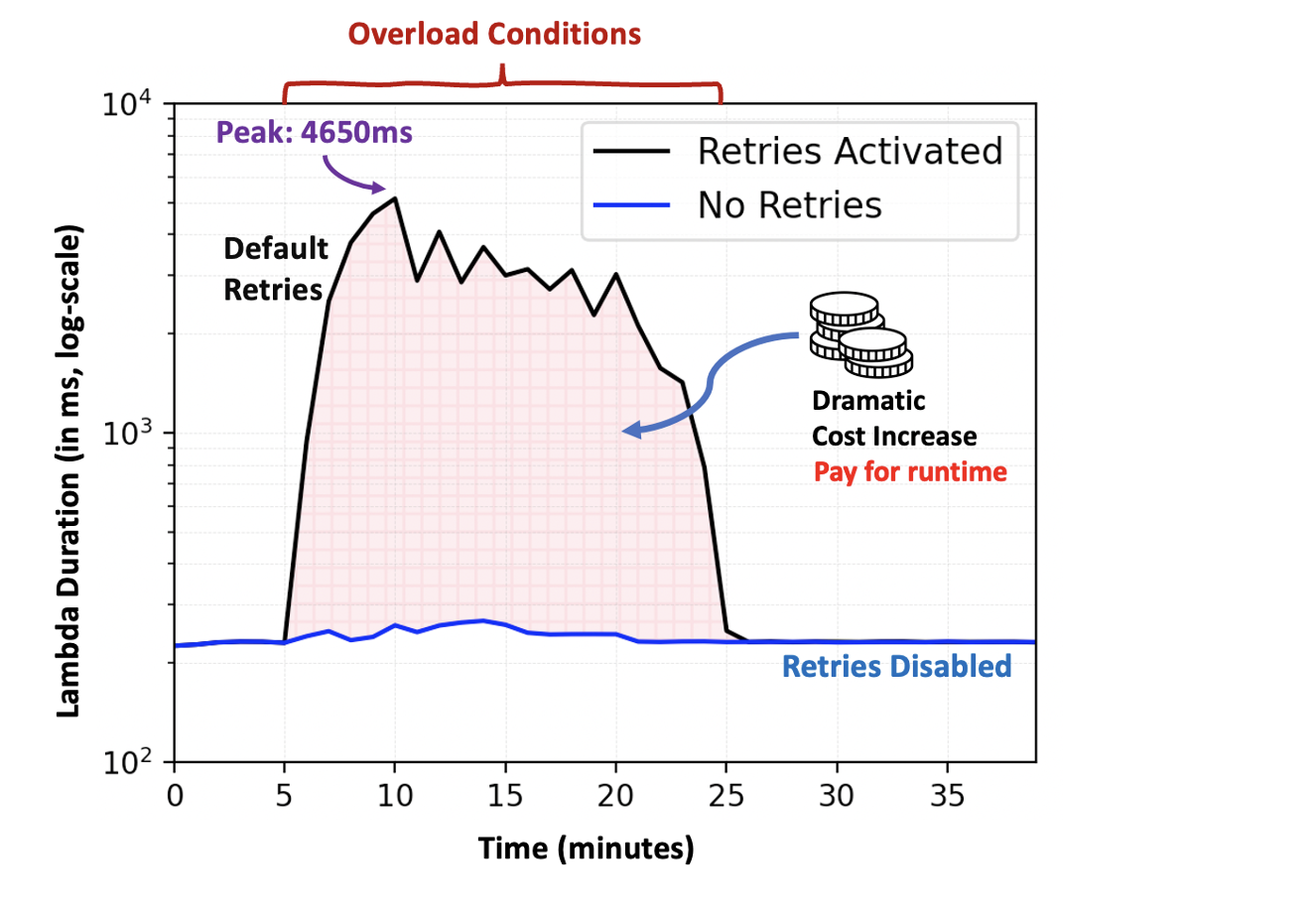}
        \caption{Lambda delay and cost (log scale): up to 2000\% latency increase during retry storms.}
        \label{fig:fig_exp_retries_lambda}
    \end{subfigure}
    \par\vspace{0.6em}
    \begin{subfigure}[c]{\linewidth}
        \centering
        \includegraphics[width=0.86\linewidth]{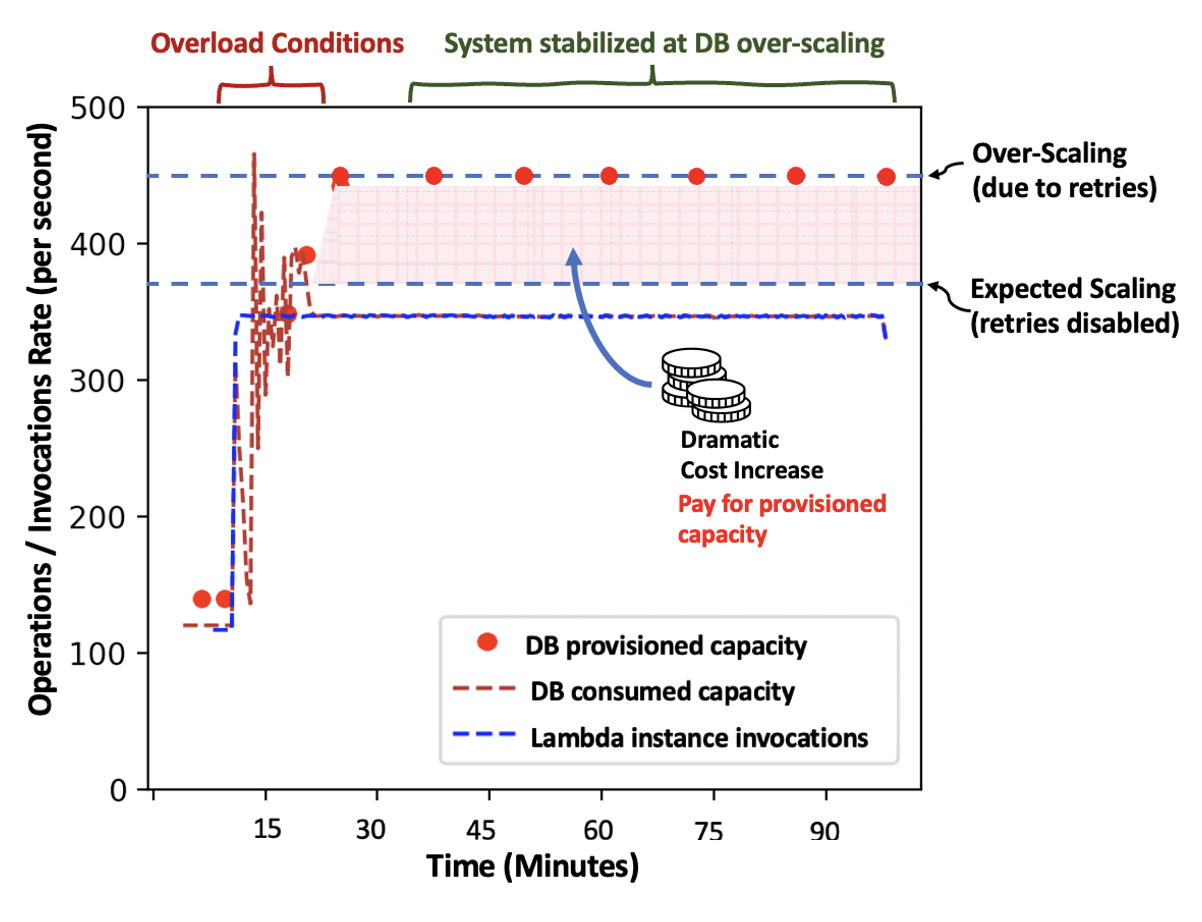}
\caption{DB over-scaling: Over-provisioned (red dots) during and after the retry storm.}
        \label{fig:fig_exp_retries_db}
    \end{subfigure}
\caption{AWS: Retry storm and Over-provisioning.}    
\label{fig:dow_aws}
\end{figure}

Measurements\footnote{We collected data on Lambda invocations and DB operations and capacity using HTTP response codes and Amazon CloudWatch reports, a monitoring service by AWS that provides metrics, alarms, and more \cite{amazoncloudwatch}.} are depicted in Figs. \ref{fig:fig_exp_retries_lambda} and \ref{fig:fig_exp_retries_db}.
Initially, the system was balanced.
Then,  at minute 5 the rate increased by 200\%.
While the Lambda function (blue curve, Fig. \ref{fig:fig_exp_retries_db}) promptly adapted to it,
DynamoDB did not\footnote{As can be seen in the red dashed curve, there is an isolated jump right away due to the 'reserved capacity' feature, which is meant to handle sudden spikes, but after a few seconds it returned to its actual provisioned size.}. DynamoDB took nearly 20 minutes to stabilize (red dotted curve, Fig. \ref{fig:fig_exp_retries_db}) with multiple scale operations.
During this period, a retry storm was triggered. Each failed DB operation forced the calling Lambda to wait and retry until success or timeout. During the DB scaling, Lambda's average latency (blue curve, Fig.~\ref{fig:fig_exp_retries_lambda}) spiked drastically 
to a 2,000\% increase compared to the typical duration.
Furthermore, due to the retry storm that overwhelmed it, the DB stabilized at 20\% over-scaling (capacity of 448 req/sec instead of the required 383 operations/sec, 90\% utilization, Fig. \ref{fig:fig_exp_retries_db}, red dots).

Repeating the experiment with retries turned off (blue curve, Fig. 4a) eliminated the excess delays, and no retry storm formed or over-scaling occurred.

\subsubsection{The Price Tag of Retry Storms}
Our experiments show that retry-based processes can substantially raise costs
in two significant ways 
even after the storm:
(1) Lambda’s costs spiked by up to 2000\% at peak load, with a substantial increase sustained throughout the 20-minute scaling period\footnote{Lambda costs are driven by invocations and latency.}.
(2) DynamoDB costs rose by about 20\% due to over-scaling\footnote{Provisioned-mode pricing depends on the provisioned capacity.}, with elevated charges potentially lasting for hours or days because of scale-down policies\footnote{AWS DynamoDB scales down only if utilization remains below a configured threshold for more than 15 consecutive minutes \cite{aws_dynamodb_docs}.}.

\subsection{Service Mesh: Istio's Bookinfo}
Having shown the challenges of retries on AWS, we now demonstrate them with the Istio Service Mesh platform \cite{istio_traffic_management}.

\subsubsection{Setup}

\begin{figure}[h]
    \centering
    \includegraphics[width=\linewidth]{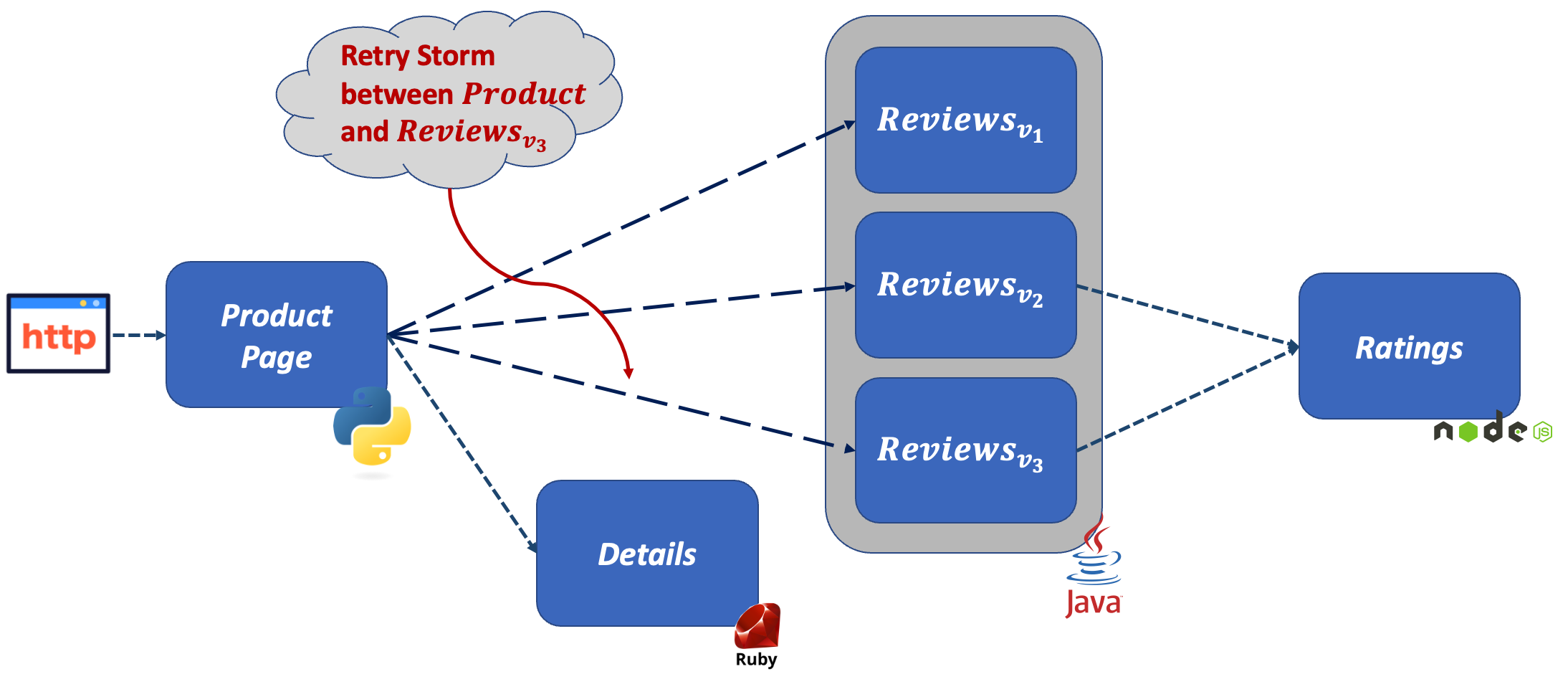}
    \caption{\textit{Istio Bookinfo}: End-to-end architecture.}
    \label{fig:bookinfo}
\end{figure}

We used the Bookinfo application~\cite{istio_bookinfo}, a widely used microservices benchmark \cite{ashok2021leveraging}, composed of four services (see Fig. \ref{fig:bookinfo}).
The 
services used Kubernetes' default autoscaling \cite{kubernetes_autoscaling}.
The \textit{Reviews} service had slower scaling decisions with a 60-second stabilization window. Istio's retry feature permitted up to five attempts. 
We collected the experimental data using Prometheus \cite{rabenstein2015prometheus}.

\subsubsection{Self-Inflicted DoW}
\begin{figure}[h]
    \centering
    \includegraphics[width=\linewidth]{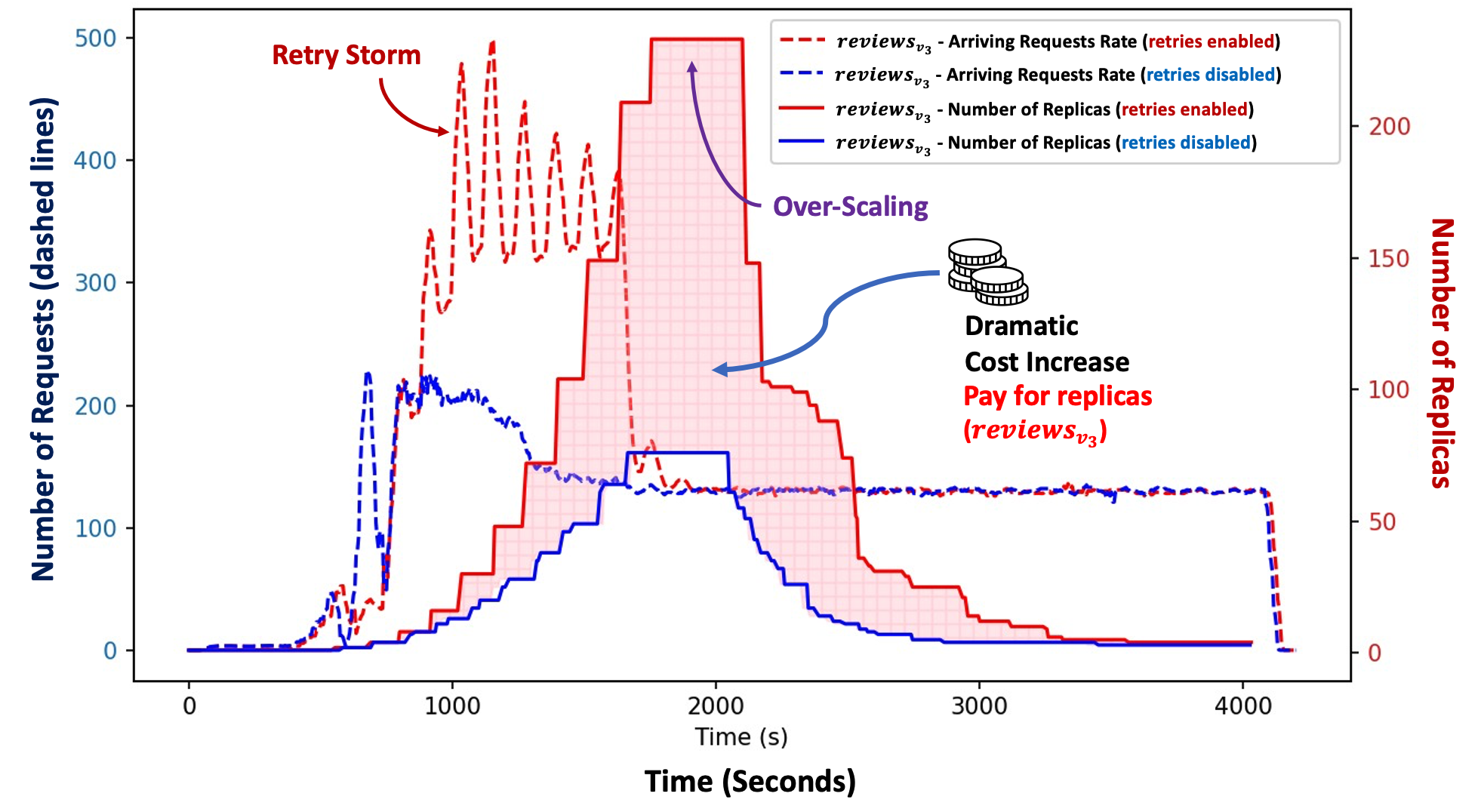}
\caption{Bookinfo: Retry storm is formed  at \textit{Product page}.}    
    \label{fig:num_pods_retries}
\end{figure}

In this experiment (Fig.~\ref{fig:num_pods_retries}), the system initially processes traffic at a stable rate. At 300 sec, the rate surges. While the \textit{Product} service scales up quickly, \textit{Reviews}, with slower scaling, lags behind. This disparity triggers a ``retry storm'' (dashed red line, Fig. \ref{fig:num_pods_retries}) as \textit{Product} continuously retries \textit{Reviews}.
Due to the retry storm, \textit{Reviews} eventually expands, inflicting significant over-charges\footnote{As with the AWS experiment, the system experiences drastic increases of  request latencies;  for the sake of brevity their analysis is omitted here.} (red continuous line, Fig. \ref{fig:num_pods_retries}) for a period of 20 minutes.

We repeated the experiment with retries disabled. \textit{Reviews} scaled to a level three times lower than its original scaling value (blue continuous line, Fig. \ref{fig:num_pods_retries}) and completed its scale-up five minutes earlier.
Furthermore, the overall rejection rate slightly decreased from 5\% to 4\%, and request latency was lower compared to the original setup by up to 30\%. 

Overall, as we observed in AWS, enabling retries during prolonged miscoordination conditions increased costs and degraded performance, offering no benefits on this test case.

\section{RetryGuard Design}\label{sec:retrydesign}

We next propose \textbf{\textit{RetryGuard}}: a mechanism
that dynamically enables or disables retries at the \textit{service level}, based on a \textit{productive-retry} controller which detects whether retries are productive (instantaneous failures) or counterproductive (miscoordination).
The design is supported by the analysis in Sec.~\ref{sec:analysis}.

\subsection{A Dynamic Controller for Productive Retries}

The productive-retry controller evaluates retry effectiveness at the service level using observable performance metrics: retry volume, request responses and error codes (rejections), and execution durations (delays). Alg.~1 presents the pseudo-code of a rejection-based variant, where \textit{measure\_value()} collects failure signals through monitoring systems such as AWS CloudWatch~\cite{amazoncloudwatch} or Google Cloud Monitoring~\cite{verginadis2023review,google_cloud_monitoring}. 
If the measured rejection rate remains above a predefined threshold for a sustained interval, retries are disabled; once it remains below the threshold, retries are re-enabled.\footnote{Alg.~1 presents the binary ON/OFF variant; the controller extends naturally to gradual re-enabling, where retries resume at a fraction of their original rate and ramp up while rejections stay below the threshold, as used in our Istio deployment (Sec.~\ref{subsec:multilayer}).}

\begin{algorithm}\label{alg:alg1}
\small
\caption{Rejection-based productive-retries controller}
\begin{algorithmic}[1]
\State Set \texttt{LowStreak} $\gets 0$,  \texttt{    HighStreak} $\gets 0$,  \texttt{Retries} $\gets$ ON
\State Set \texttt{Threshold} (parameter) and \texttt{Interval} (parameter)

\While{true}
    \State \texttt{Failures} $\gets$ \texttt{measure\_value()}
    \If{\texttt{Failures} $<$ \texttt{Threshold}}
        \State \texttt{LowStreak} $\gets$ \texttt{LowStreak} $+ 1$
        \State \texttt{    HighStreak} $\gets 0$
    \ElsIf{\texttt{Failures} $>$ \texttt{Threshold}}
        \State \texttt{    HighStreak} $\gets$ \texttt{    HighStreak} $+ 1$
        \State \texttt{LowStreak} $\gets 0$
    \Else
        \State \texttt{LowStreak} $\gets 0$ and \texttt{    HighStreak} $\gets 0$
    \EndIf

    \If{\texttt{LowStreak} $\geq \text{Interval}$}
         \texttt{Retries} $\gets$ ON
    \ElsIf{\texttt{    HighStreak} $\geq \text{Interval}$}
         \texttt{Retries} $\gets$ OFF
    \EndIf
\EndWhile
\end{algorithmic}
\end{algorithm}

Note that \textit{RetryGuard}'s actions are active only during prolonged periods. 
This service-layer operation reflects a philosophy similar to BBR \cite{cardwell2016bbr}, which proposes to proactively adjust traffic at the transport layer.
Thus, in well-scaled or instantaneous errors, \textit{RetryGuard} matches the performance of legacy mechanisms, while during miscoordination, it outperforms them by preemptively preventing nonproductive attempts.

\subsection{Comparison to Common Retry Mechanisms}

\begin{figure*}[t]
    \centering
    \begin{subfigure}{0.24\linewidth}
        \centering
        \includegraphics[width=\linewidth]{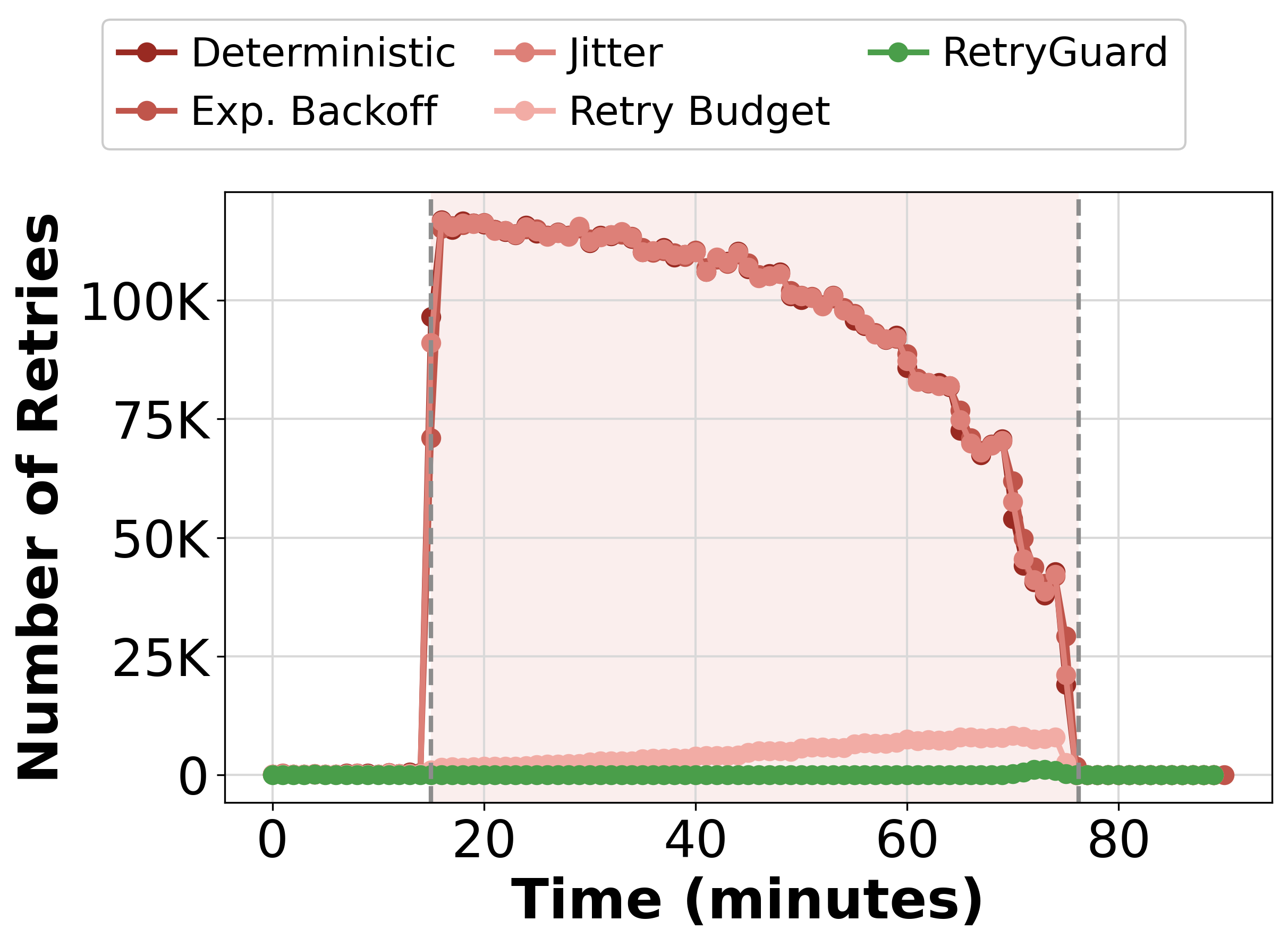}
        \caption{Number of retries.}
        \label{fig:sim_storm}
    \end{subfigure}
    \hfill
    \begin{subfigure}{0.24\linewidth}
        \centering
        \includegraphics[width=\linewidth]{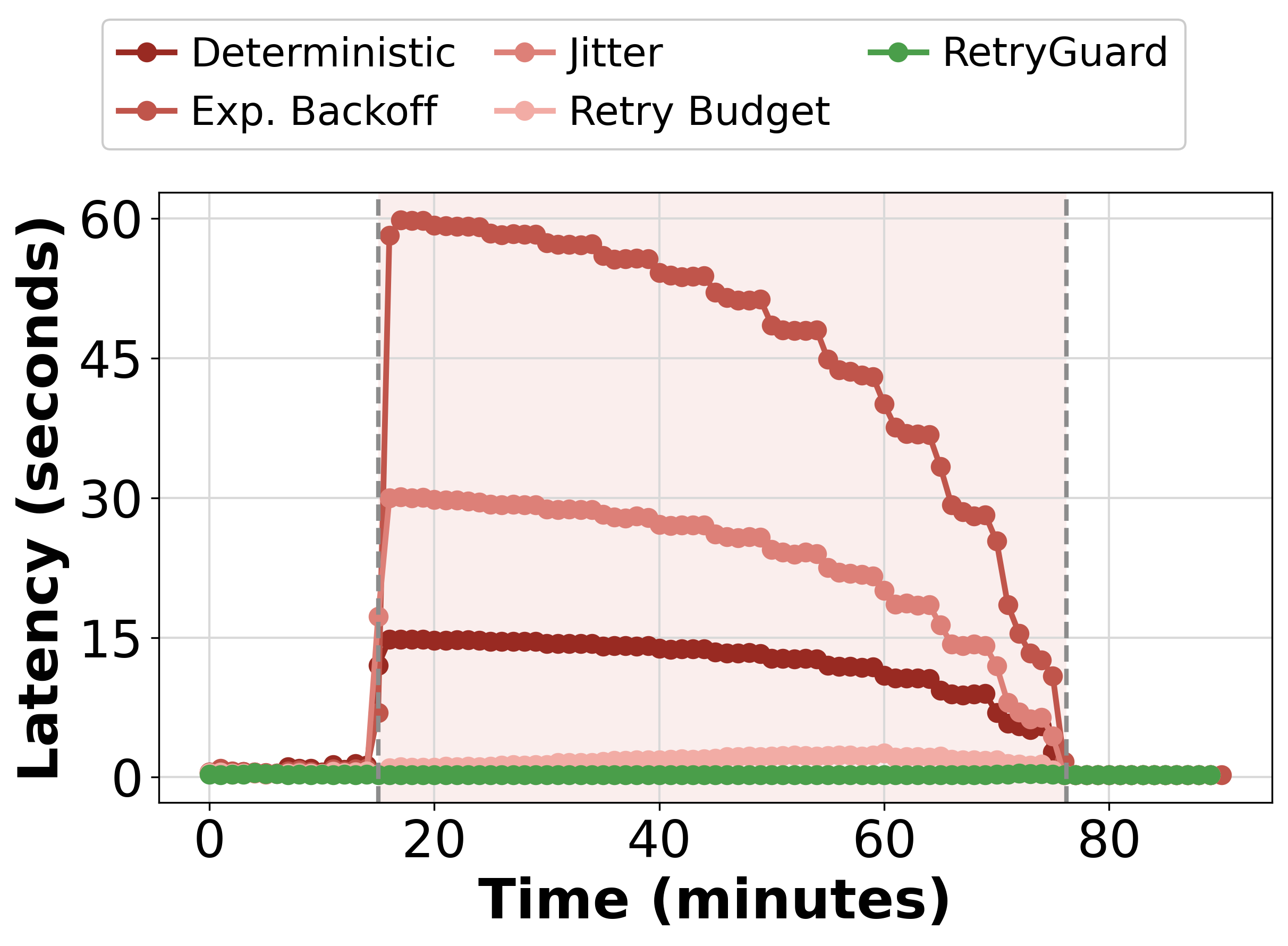}
        \caption{Request latency.}
        \label{fig:sim_duration}
    \end{subfigure}
    \hfill
    \begin{subfigure}{0.24\linewidth}
        \centering
        \includegraphics[width=\linewidth]{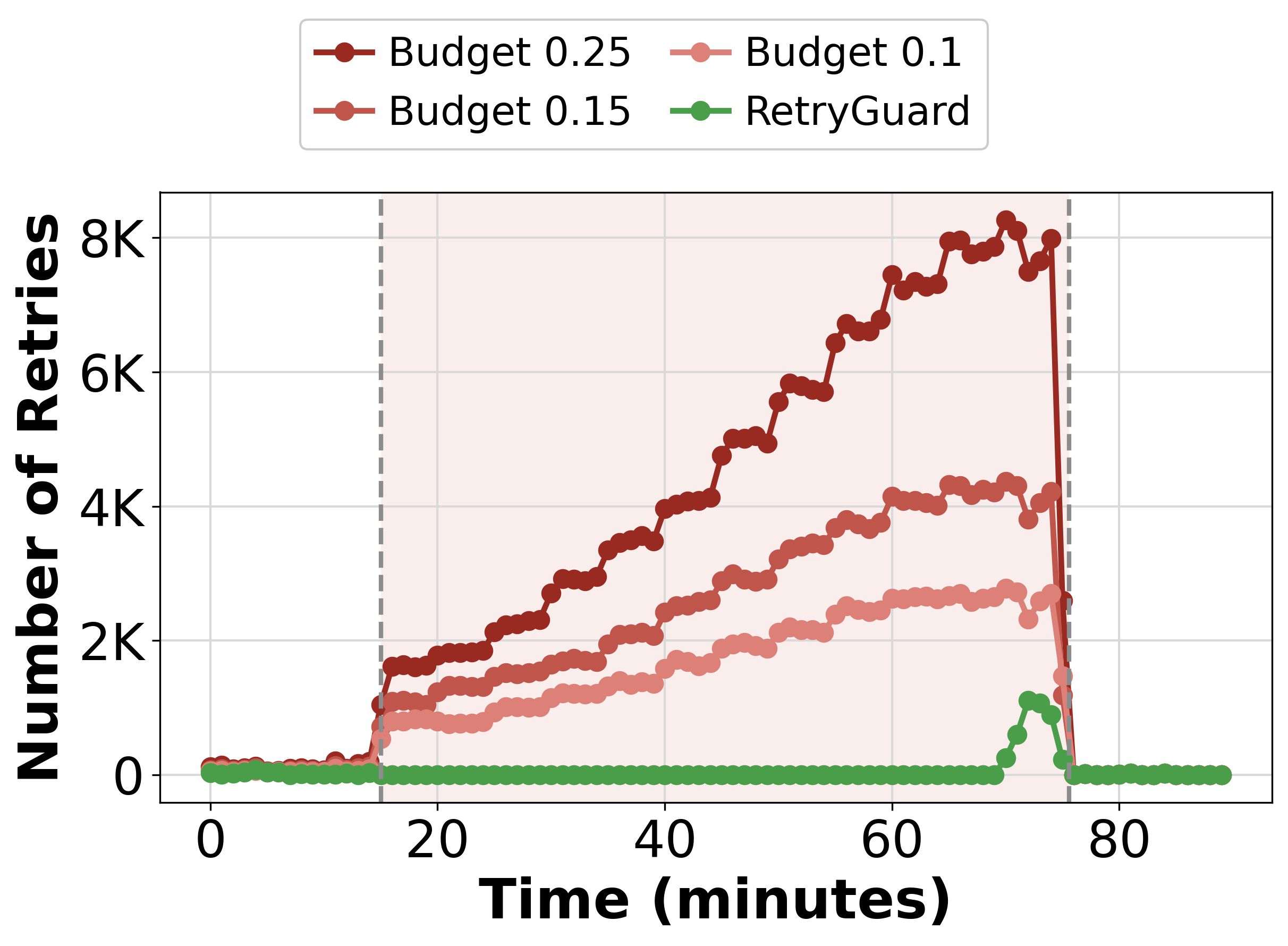}
        \caption{Number of retries.}
        \label{fig:sim_budget_storm}
    \end{subfigure}
    \hfill
    \begin{subfigure}{0.24\linewidth}
        \centering
        \includegraphics[width=\linewidth]{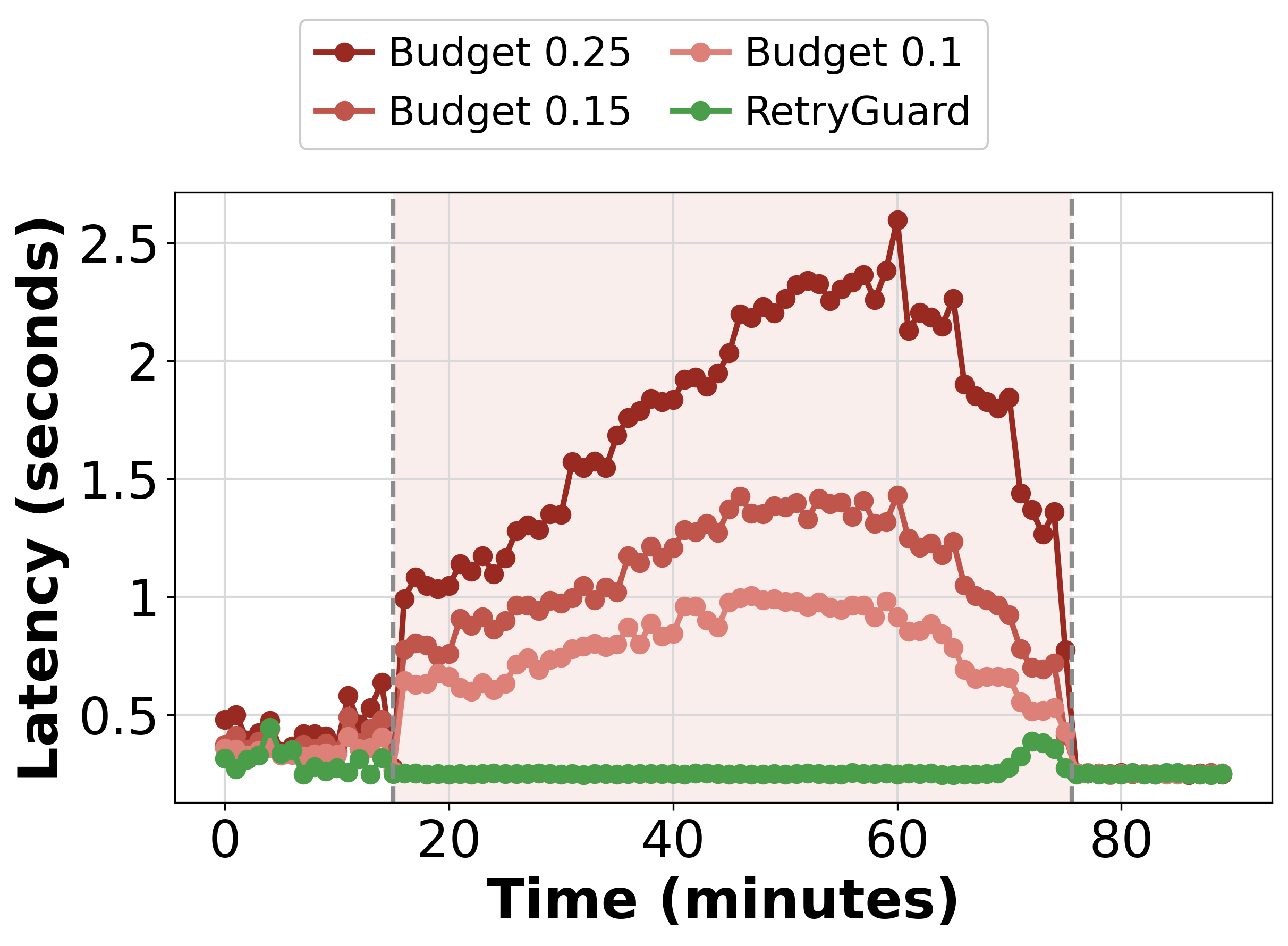}
        \caption{Request latency.}
        \label{fig:sim_budget_duration}
    \end{subfigure}
\caption{\textit{RetryGuard} compared with deterministic timeouts, exponential backoff, jitter, and retry budgets. Panels (a) and (b) compare all mechanisms, while panels (c) and (d) zoom in on retry budgets under several parameter settings.}
\label{fig:sim}
\end{figure*}

Conventional retry strategies such as exponential backoff and jitter primarily control retry timing by delaying or randomizing subsequent attempts. While effective for transient failures, under sustained miscoordination they only spread redundant retries over time and may substantially increase request latency. As shown in Fig.~\ref{fig:sim}(a) and (b), both mechanisms continue generating large retry volumes and high latency throughout the overload period.

Retry budgets appear more suitable because they explicitly limit retry volume. However, they still permit retries during prolonged overload, even when such attempts are unlikely to succeed. Figures~\ref{fig:sim}(a) and (b) show that retry budgets reduce amplification relative to backoff and jitter, while panels (c) and (d) examine several budget values in greater detail. Reducing the budget (0.25 to 0.1) consistently lowers retry volume and latency, but does not eliminate redundant attempts. \textit{RetryGuard} instead detects the sustained-overload regime and disables counterproductive retries, except for brief recovery probes, preserving downstream capacity for first attempts.

\subsection{RetryGuard Principles, Properties and Parameters}
\label{subsec:retryguaed_prinicples}

\subsubsection{Distributed, Minimal Coordination}

\textit{RetryGuard} requires no coordination between the source and destination services, making it ideal for distributed deployment across multiple points in a microservices graph, allowing each service to manage retries locally without the need for a centralized orchestrator.
It can be implemented solely at the source  without altering downstream services. Specific implementations may leverage shared observability tools or centralized metric systems (e.g., Prometheus \cite{rabenstein2015prometheus} or AWS CloudWatch \cite{amazoncloudwatch}).

\subsubsection{Seamless Operation in Coordinated (Balanced) Systems}
Importantly, \textit{RetryGuard} is designed to be \textbf{non-disruptive} during stability conditions, ensuring seamless operation when service traffic is well-balanced and scalable. 
If service traffic is well-balanced, the performance metrics will not exceed the threshold (such probabilities are negligible, as analyzed in Sec. \ref{sec:analysis}). Thus, a sudden traffic burst will not activate intervention, and whenever a system is ``ideal'' or  ``perfectly scaled'', \textit{RetryGuard} remains non-intrusive.
\\

\noindent
Overall, \textit{RetryGuard} acts as a safeguard that prevents clients from compounding a prolonged period of miscoordination and overload with excessive retry attempts, while keeping the benefits of the classic retry mechanism in normal operation.

\subsection{RetryGuard Implementation}\label{subsec:retryguard_implementation}

\subsubsection{Surrogate Threshold Metrics}
 \textit{RetryGuard} can rely on several metrics, depending on their availability. If retry volumes per request are measurable (e.g., through retry counters), \textit{RetryGuard} can track it to identify unproductive retries. If retry metrics are not accessible, we can instead rely on rejection signals (e.g., error logs or rejection codes). If neither is available, \textit{RetryGuard} can use response delays (monitored via tools like AWS CloudWatch \cite{amazoncloudwatch}) as indicators, temporarily halting retries during high latency and reintroducing them as delays normalize.  \textit{RetryGuard} is evaluated using rejection rate and response delays, to demonstrate its flexibility.\footnote{
Sec.~\ref{sec:analysis} introduces a mathematical framework that captures the interplay between retries, throughput and delays, offering practical threshold guidance.}

\subsubsection{RetryGuard Parameter Tradeoff}
\label{subsec:streak_sensitivity}
\textit{RetryGuard} requires two parameters: an activation threshold and low/high streak durations for disabling and re-enabling retries. In the setup of Sec.~\ref{sec:evaluation}, shorter streaks reduced unnecessary retries under light overload by over $90\%$, but increased the stable-load rejection rate from $1.0\%$ to $5.3\%$. Under extreme DDoS load, both settings yielded similar rejection rates as performance was dominated by downstream capacity. Sec.~\ref{sec:analysis} shows that the sharp retry-rate transition near $\rho=1$ makes the threshold robust to configure. Both parameters remain deployment-specific and were adapted to each system's scaling dynamics in Sec.~\ref{sec:evaluation}.

\subsubsection{Integrating RetryGuard}
\textit{RetryGuard} can be deployed in both closed-source (e.g., AWS) and open-source (e.g., Istio) environments by leveraging existing observability signals, enabling adaptive retry control without modifying application code or existing request paths. In AWS, we deployed \textit{RetryGuard} by continuously monitoring services on a dedicated EC2 instance via existing telemetry (CloudWatch) and dynamically updating service configurations, so its decisions neither block nor delay user traffic 
in-flight.
In Istio, \textit{RetryGuard} was implemented as a dedicated pod that periodically samples Istio’s metrics interface, imposing no per-request overhead.

\section{Analysis: Load, Delay and Cost}
\label{sec:analysis}
In this section we use an analytical model to examine the critical transition of a service from stable-load to miscoordination and to support \textit{RetryGuard} in two key aspects: (1) In establishing key design principles and in deriving threshold activation guidelines for the Productive-Retry Controller; and (2) In estimating the costs of retries under miscoordination, as well as retry rate, rejection probability, and expected delay.

In Sec. \ref{subsection:part1}, 
we formulate the mathematical framework and analyze it under miscoordination. Sec. \ref{subsection:stableload} analyzes a service under stable-load conditions. In Sec. \ref{subsec:criticalpoint}, we examine the stable-load/miscoordination transition and evaluate \textit{RetryGuard}’s robustness. Our analysis establishes that
\textit{RetryGuard} can robustly detect miscoordination (overload) while remaining non-intrusive under stable-load.

\subsection{Retries and Rejections Under Miscoordination}\label{subsection:part1}

\begin{figure}[t]
    \centering
    \includegraphics[width=\linewidth]{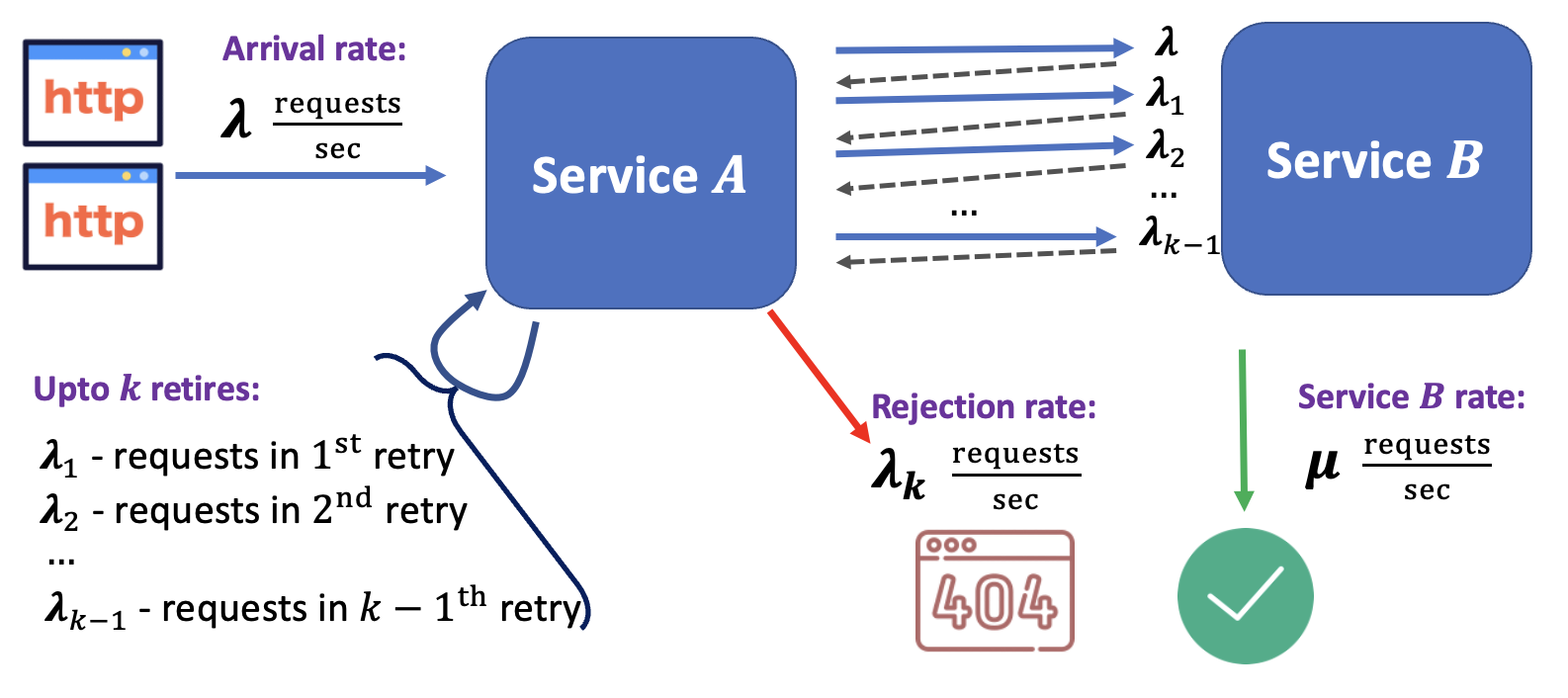}
    \caption{The Model and notation.}
    \label{fig:analyysisnotation}
\end{figure}

The system under consideration consists of Service A and Service B operating in tandem (see Fig. \ref{fig:analyysisnotation}). Service $A$ receives incoming requests at  rate $\lambda$ (reqs/sec). Service A holds the transactions and attempts to forward them to Service B. Let $\mu$ denote the rate at which Service B can process requests (i.e., its capacity  in requests per second). If Service B cannot accept the requests due to capacity constraints, Service A will retry later. We assume that Service A imposes a limit on the number of retries per request, denoted by $k$. Let $\lambda_i$ represent the rate of requests currently in their $i$-th retry attempt, where $i = 1, \dots, k$. This notation is illustrated in Fig. \ref{fig:analyysisnotation}.

We analyze the steady-state behavior of the system in a miscoordination scenario, where $\rho = \lambda / \mu > 1$, using a fluid approximation of aggregate retry traffic under sustained, nearly stationary, overload. The model captures average request and retry rates rather than per-request queueing dynamics or short-timescale correlations. Fig.~\ref{fig:criticaltransition} validates it against finite-buffer $M/M/1/m$ simulations across stable and overloaded regimes. 
We define $p$ as the success probability of an arbitrary request by \textit{Service B}\footnote{Rejection indicates that, at that specific attempt, \textit{Service B} could not process the request. During the first $k-1$ attempts, the request will retry according to the defined retry pattern. However, on the $k$th attempt, the request will be considered ``failed'' (i.e., throttled) by \textit{Service B}, as illustrated by the red arrow in Figure \ref{fig:analyysisnotation}.}. Let $\tilde{p} := 1 - p$ be the rejection probability. Note that  $p$  depends on the  retries rate, as we show in Eq. \eqref{eq:pdef} below.

\subsubsection{Deriving $\lambda_i$ and $\tilde{p}$}

The total volume of requests that \textit{Service B} receives per second is: 
\begin{equation}\label{eq:lambdasum}
    \Lambda = \lambda + \textstyle \sum_{i=1}^{k-1} \lambda_i,
\end{equation}  
which includes new requests, $\lambda$, and retries, $\lambda_i$.
Since the capacity of \textit{Service B} is $\mu$, and $\mu < \lambda < \Lambda$ in an overload scenario, the success probability of an arbitrary request, $p$, and, consequently, the rejection probability $\tilde p$,  are given by:  
\begin{equation}\label{eq:pdef}
    p = \nicefrac{\mu}{\Lambda} = \nicefrac{\mu}{\left( \lambda + \sum_{i=1}^{k-1} \lambda_i \right)}; 
    ~~~~~~~
     \tilde{p} = 1 - \nicefrac{\mu}{\Lambda}.
\end{equation}
Under the fluid approximation, all attempt classes share the same downstream capacity and observe the same aggregate congestion state. We therefore approximate their time-averaged rejection fractions by a common value $\tilde p$.
Under this assumption:  $
    \lambda_1 = \tilde{p} \cdot \lambda$ and $\lambda_i = \tilde{p} \cdot \lambda_{i-1}$ for $i = 2, \dots, k$.
Expanding this relationship yields  
    $\lambda_i = \tilde{p}^i \cdot \lambda, \quad \text{for } i = 1, \dots, k$.
By substituting this expression into Eq. \eqref{eq:pdef} and Eq. (\ref{eq:lambdasum}), and through algebraic simplifications 
(due to lack of space, we omit the intermediate steps)
we derive:
\begin{equation}\label{eq:pvalue}
    \tilde{p} = \sqrt[k]{1 - \nicefrac{\mu}{\lambda}} \quad \text{and} \quad     \lambda_i \approx \lambda \cdot \left( 1 - \nicefrac{\mu}{\lambda} \right)^{\frac{i}{k}}.
\end{equation}  

\begin{figure}[t]
    \centering
    \includegraphics[width=\linewidth]{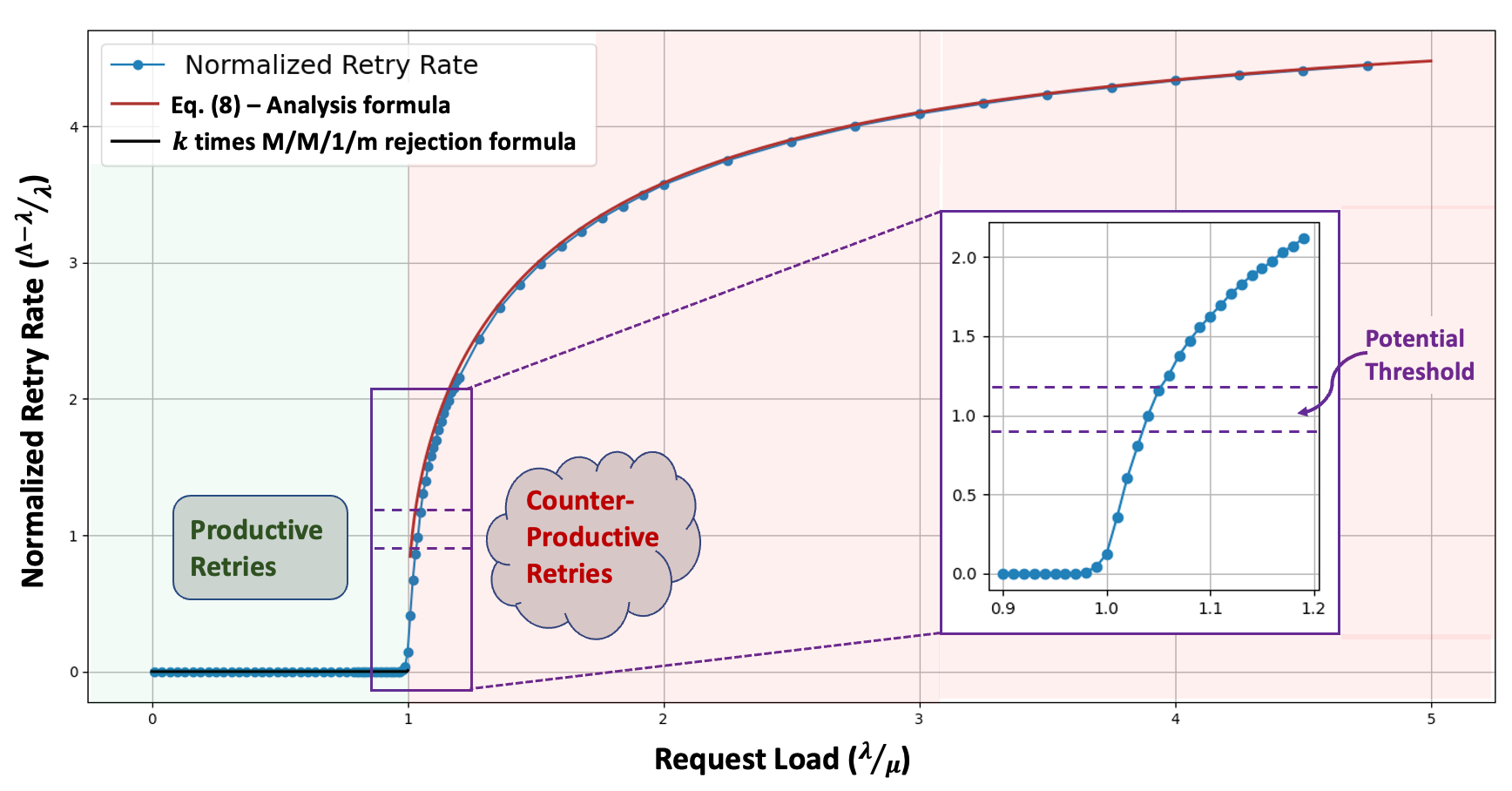}
    \caption{Normalized retry rate as a function of request load. The zoomed-in panel focuses on the critical transition range where retries rise sharply.}
    \label{fig:criticaltransition}
\end{figure}

\begin{remark}
The accuracy of the analysis is verified by a comparison to simulation in Fig. \ref{fig:criticaltransition}. Examining the curves to the right of load = 1, where the red curve represents the approximation and the blue dots represent simulation results (on M/M/1/m), shows very good agreement. 
\end{remark} 

\subsubsection{Deriving the Expected Delay $E[T]$}

Next, we derive the expected delay (or lifetime) of a request in the system, focusing on the well-known Exponential Backoff strategy for retries. While our analysis centers on Exponential Backoff, similar methods can be applied to other retry strategies, such as Jitter or Linear Increasing Delays.
Retries follow a truncated geometric distribution. Let $X$ represent the number of retries experienced by an arbitrary request. Then:
\begin{equation}
\label{eq:X}
  Pr[X = i] =
    \begin{cases}
      (1 - p)^i \cdot p & 0 \le i \le k - 1, \\
      (1 - p)^k & i = k
    \end{cases}       
\end{equation}
and $0$ for any $i > k$.
In Exponential Backoff, the waiting time between retries increases with each attempt, following powers of 2. Let $T_i$ denote the total delay experienced by a request with $i$ retries:
$T_i = 1 + 2 + \dots + 2^{i-1} = 2^i - 1$.
Using Eq.~\eqref{eq:X}, the expected delay of a request is:
$
E[T] = \sum_{i=0}^k Pr[X = i] \cdot T_i
$.
After algebraic manipulation, we obtain:
\begin{equation}
E[T] =
\frac {2 \tilde{p} p [ (2 \tilde{p})^k - 1 ]}{2 \tilde{p} - 1}
-
\frac {\tilde{p} p [ \tilde{p}^k - 1 ] } {\tilde{p} - 1 }
+ \tilde{p}^k (2^k - 1).
\end{equation}

This metric can be used to estimate the costs incurred by Service A on the applications, since they are proportional to the time spent by a request in the system (i.e., its delay).

\subsection{Behavior under Stability Conditions}
\label{subsection:stableload}

To ensure that \textit{RetryGuard} remains non-intrusive under stable-load conditions, we analyze its behavior using the M/M/1/m queueing model, which is widely known in the literature for modeling systems with finite buffer capacity. In this model, $\lambda < \mu$, ensuring a system under stability conditions. The rejection probability of a request, $\tilde{p}$, can be expressed as:
\begin {equation}
\tilde{p} = 
\nicefrac {(1- \rho) \rho^m}
{1-\rho^{m+1}}
\label{eq:stable-load}
\end {equation}
where $m$ is the buffer capacity of Service B and $\rho = \lambda/\mu$.

This allows us to estimate the likelihood for a request that is submitted to Service B  to be rejected and retried under stability conditions. Fig. \ref{fig:criticaltransition}  (to the left of $\rho = 1$) depicts the retry rate (black line) based on this analysis and estimated at $k \tilde p$, closely matching simulation results (blue dots).

\begin{figure*}[t]
    \centering
    \begin{subfigure}{0.225\linewidth}
        \centering
        \includegraphics[width=\linewidth]{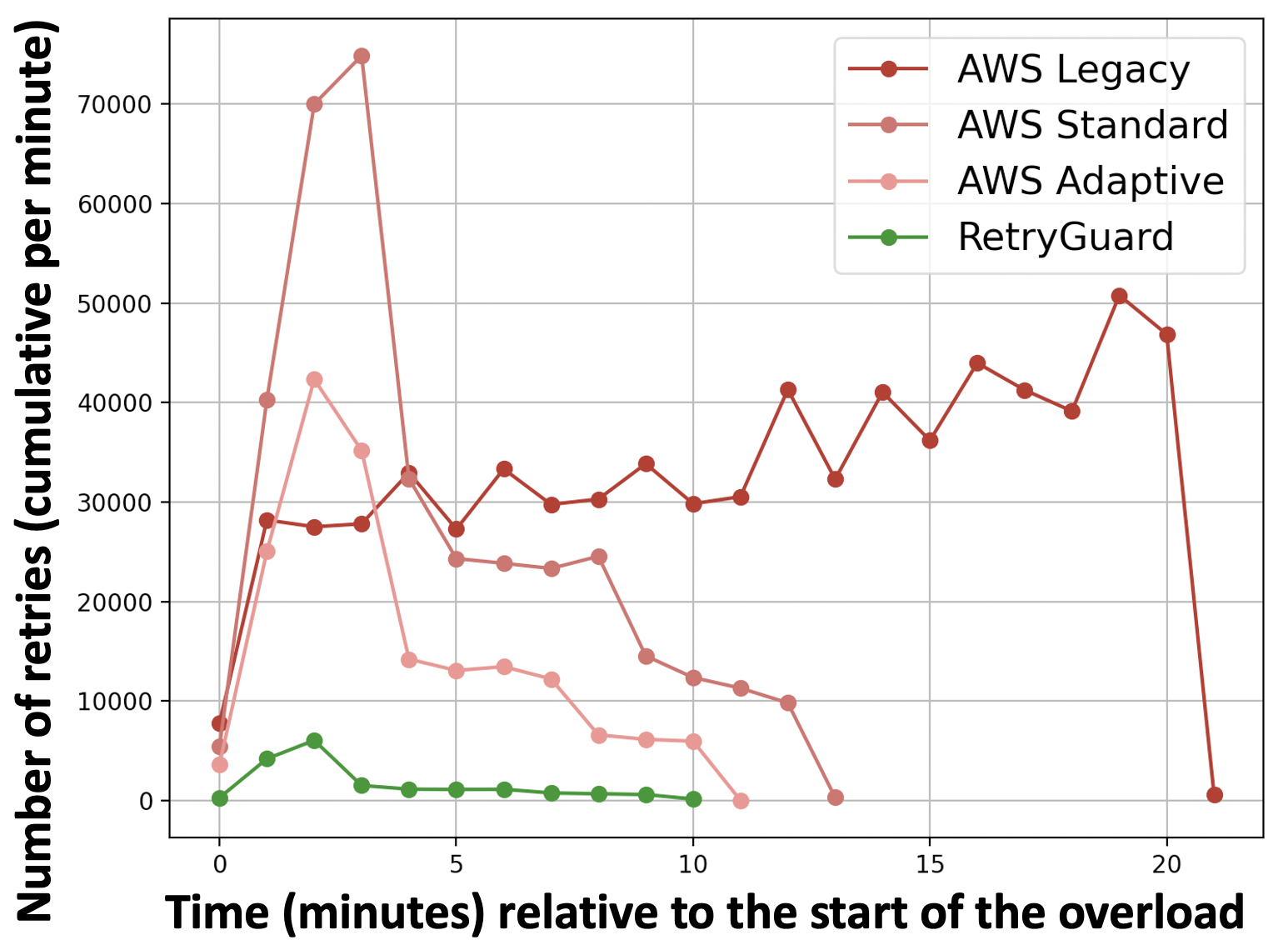}
        \caption{Number of retries.}
        \label{fig:storm}
    \end{subfigure}
    \hfill
    \begin{subfigure}{0.235\linewidth}
        \centering
        \includegraphics[width=\linewidth]{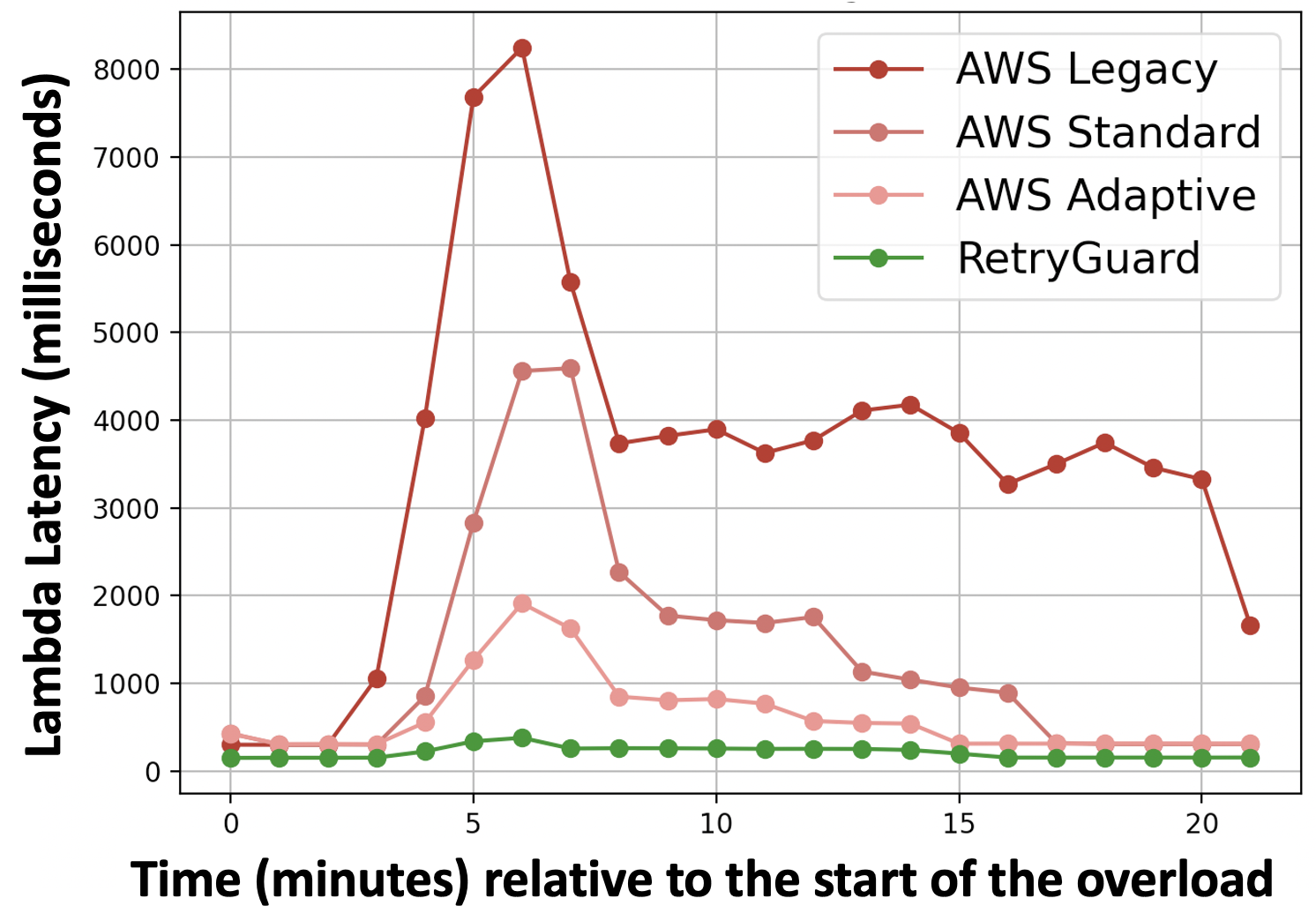}
        \caption{Average Lambda Latency}
        \label{fig:duration}
    \end{subfigure}
    \hfill
    \begin{subfigure}{0.25\linewidth}
        \centering
        \includegraphics[width=\linewidth]{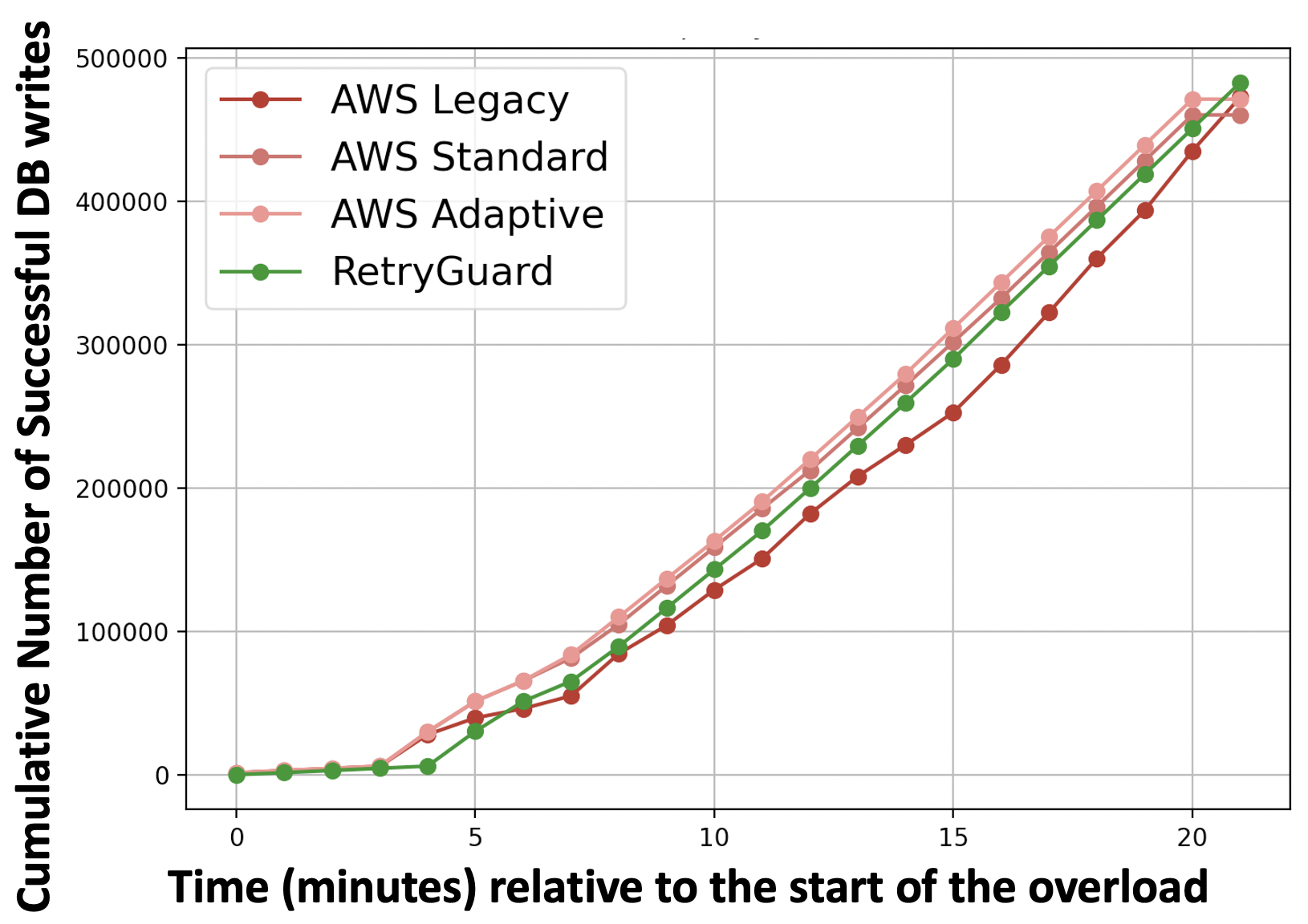}
        \caption{Cumulative DB Writes}
        \label{fig:writes}
    \end{subfigure}
    \hfill
    \begin{subfigure}{0.25\linewidth}
        \centering
        \includegraphics[width=\linewidth]{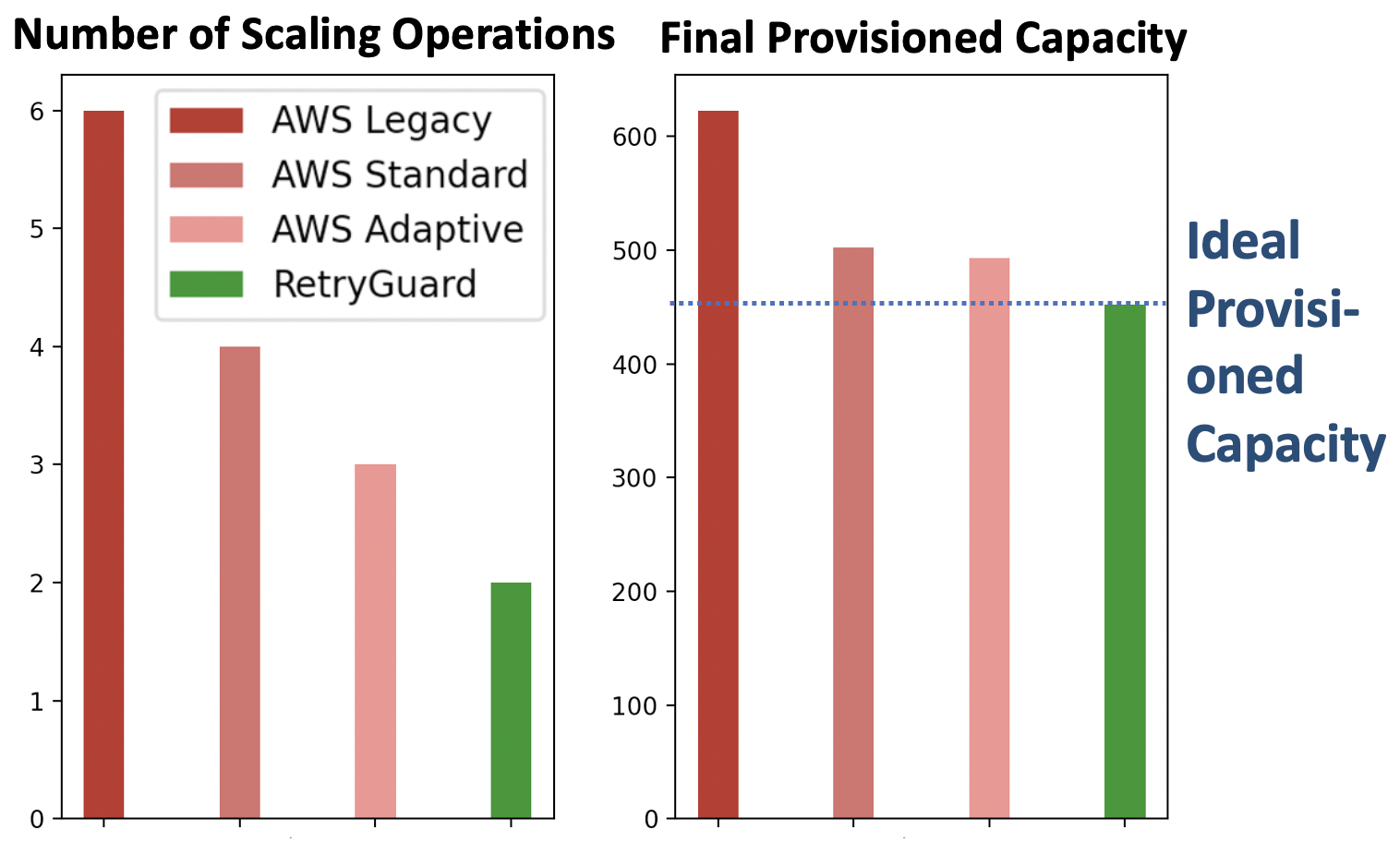}
        \caption{DynamoDB Scaling Stats}
        \label{fig:dbscaling}
    \end{subfigure}
\caption{\textit{RetryGuard} achieves lower retry rate and latency while maintaining the rejection rate (DB writes).}
    \label{fig:exp_aws}
\end{figure*}

\subsection{Practical and Operational Implications}\label{subsec:criticalpoint}

\subsubsection{\textbf{RetryGuard's transitions between stability conditions and miscoordination are robust}}
\label{sec:5.3.1}
The analysis provided above (Eqs. (\ref{eq:pvalue}) and (\ref{eq:stable-load})) establishes that the retry rate undergoes a very sharp increase as the system moves from stability conditions  to miscoordination conditions. 
Fig. \ref{fig:criticaltransition} demonstrates this phenomenon by depicting the normalized retry rate as a function of the request load (load due to original requests). The reader may observe that a sharp increase (steep derivative) at $\rho =1$ marks the transition point from stable-load to over-load. 

This sharp increase allows  \textit{RetryGuard} to robustly identify the transition threshold from stability conditions to miscoordination and vice versa, since simple counting of retries (or other related parameters) over a short period of time  will yield a highly robust identifier. For example,  one can measure the normalized retry rate and select any value between 0.95 and 1.1  (see purple-dashed lines in Fig. \ref{fig:criticaltransition}) as a transition threshold value. Due to the sharp increase of the curve, this selection is robust to both the specific threshold chosen and measurement noise.
The sharp increase phenomenon occurs since at $\rho <1$ retries occur only at rare events of blocking due to statistical variation of the arrival process. In contrast, at $\rho > 1$ at least a fraction $\rho -1$ of the 
offered load is blocked, and the resulting retry traffic is amplified by a factor of $k$ due to repeated retries.

\subsubsection{\textbf{The Price Tag of The Storm}}
Evaluating $E[T]$ from Eq.~(5) across the system parameters $\lambda$, $\mu$, and $k$ reveals two key insights.
First, delay and cost grow rapidly with the retry limit $k$: as $\tilde{p}\!\to\!1$, exponential backoff makes the expected request lifetime exponential in $k$. Second, even 'low' or 'mild' overload ($\rho$ slightly above $1$) causes substantial cost increases as retries accumulate. Sec.~\ref{sec:evaluation} confirms these analytical insights through measured latency and billing.

\section{Evaluation}\label{sec:evaluation}
We evaluate \textit{RetryGuard} to demonstrate its practicality, effectiveness, and advantages over existing retry patterns. 
The evaluation is organized as follows: 
In Sec. \ref{subsec:experiments}, we deploy and compare \textit{RetryGuard} to AWS's Standard, Legacy, and Adaptive retry policies \cite{boto3-retries}. This comparison also covers the fundamental retry patterns exponential backoff, jitter, and token-bucket.
In Sec. \ref{subsec:multilayer}, we deploy \textit{RetryGuard} in a multi-layer distributed application using the Kubernetes Istio Bookinfo environment, showing its improved results.
Table \ref{tab:retryguard_comparison_combined} (Page 2) provides an overview of the main results.

\subsection{RetryGuard and AWS Retries}
\label{subsec:experiments}

We conducted several experiments using the AWS application from Sec. \ref{sec:3}.
\emph{RetryGuard} was implemented as described in Sec. \ref{subsec:retryguard_implementation}. 
That is, the AWS Gateway-Lambda-DynamoDB application was deployed, traffic was launched at it using an EC2 instance \cite{ec2_autoscaling}, and metrics were measured with CloudWatch \cite{amazoncloudwatch}. Each experiment was repeated three times.

We compared the performance of \emph{RetryGuard} against the AWS Boto3\footnote{The official AWS Software Development Kit for Python.} retry patterns \cite{boto3-retries}:
\begin{itemize}[leftmargin=*]
    \item \textbf{Legacy} (default in Boto3) uses attempts with backoff, continuing to retry under most network-related exceptions.
    \item \textbf{Standard} applies exponential backoff with jitter, respecting maximum number of attempts and delay constraints.
    \item \textbf{Adaptive} uses a token-bucket approach, relying on client-side observed throughput to limit the retry rate.
\end{itemize}

We measured several key metrics to assess both cost and performance under miscoordination: 
(1) \textbf{Retry Storm Size} (Fig. \ref{fig:storm}): the  number of retry attempts triggered.
(2) \textbf{Lambda Latency} (Fig. \ref{fig:duration}): the average delay experienced by Lambda requests throughout the experiment. 
(3) \textbf{Number of Successful Requests} (Fig. \ref{fig:writes}): the number of requests ultimately succeeded (whether on the first attempt or after retries), reflecting overall throughput. 
(4) \textbf{Number of Scaling Operations} (Fig. \ref{fig:dbscaling}, left): the number of times DynamoDB scaled up. 
(5) \textbf{Provisioned Capacity} (Fig. \ref{fig:dbscaling}, right): the final capacity provisioned by the DynamoDB, which directly impacts cost.

The measurements begin when traffic increases and miscoordination occurs, and continue until the system scales and reaches a stable state.
Fig. \ref{fig:storm} shows that \emph{RetryGuard} substantially reduces the size of the retry storm, eliminating many unnecessary retry attempts. Specifically, \emph{RetryGuard} lowered the average number of retry attempts per request from 2.09 (with the default Legacy pattern) to only 0.05, representing a 98\% decrease.
Subsequently, as shown in Fig. \ref{fig:duration}, \emph{RetryGuard} achieves significantly lower end-to-end latency than all other AWS retry patterns, resulting in substantial cost savings ranging from 40\% to 90\% just for the Lambda billing.

Fig. \ref{fig:dbscaling} illustrates that \emph{RetryGuard} minimized scaling costs while stabilizing the system. It reduces the number of  scaling operations, and eliminates over-scaling, as without it DynamoDB’s provisioned capacity exceeds actual needs.

Despite these improvements, Fig. \ref{fig:writes} shows that \emph{RetryGuard} achieves a similar number of successful writes to the DB compared to all other retry patterns. Naturally, a portion of requests fail due to DynamoDB overload, but additional retries beyond saturation do not improve effective throughput, resulting in a similar overall rejection rate for all patterns. Hence, while \emph{RetryGuard} delivers the best performance, it also yields the lowest costs.

\subsection{Multi-layer Distributed Deployment}
\label{subsec:multilayer}

We evaluate \emph{RetryGuard} in a multi-layer environment using \emph{Bookinfo} (Sec. \ref{sec:3}). 
We deployed \emph{RetryGuard} in a distributed manner, running independently at each microservice. The deployment leverages the service mesh framework, which exposes the rejection rate of each service.
If the rejection rate exceeds the threshold (20\% in the experiment) for an interval of 30 seconds, retries are disabled for every request at that service; once it drops below the threshold, retries are \textit{gradually} re-enabled again. We use the same experimental setup, metrics, and load-generation method described in Sec. \ref{sec:3}.

\begin{figure}[t]
    \centering
    \begin{subfigure}{0.46\linewidth}
        \centering
        \includegraphics[width=\linewidth]{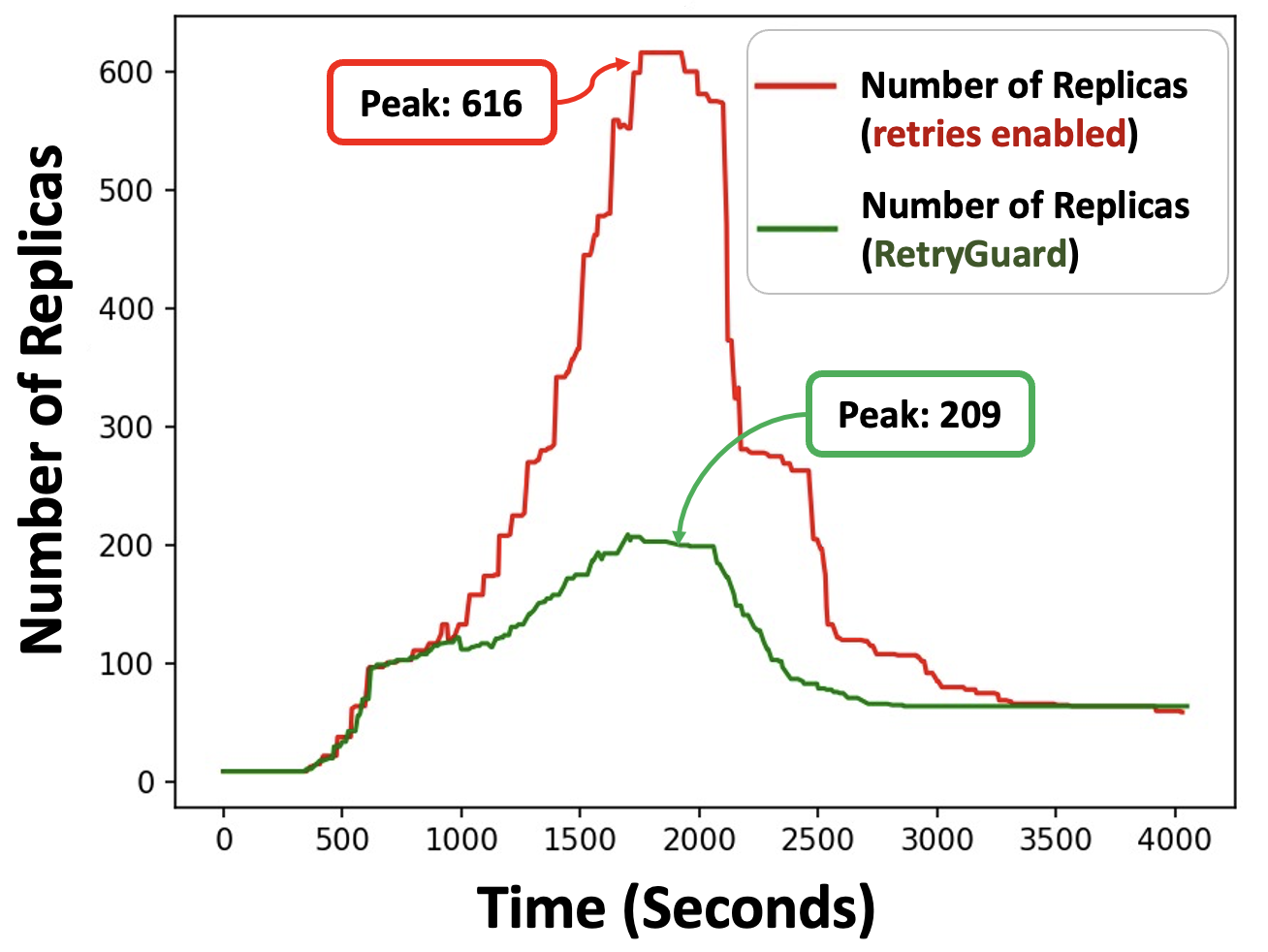}
        \caption{Total number of replicas.}
        \label{fig:num_pods_solution}
    \end{subfigure}
    \hfill
    \begin{subfigure}{0.49\linewidth}
        \centering
        \includegraphics[width=\linewidth]{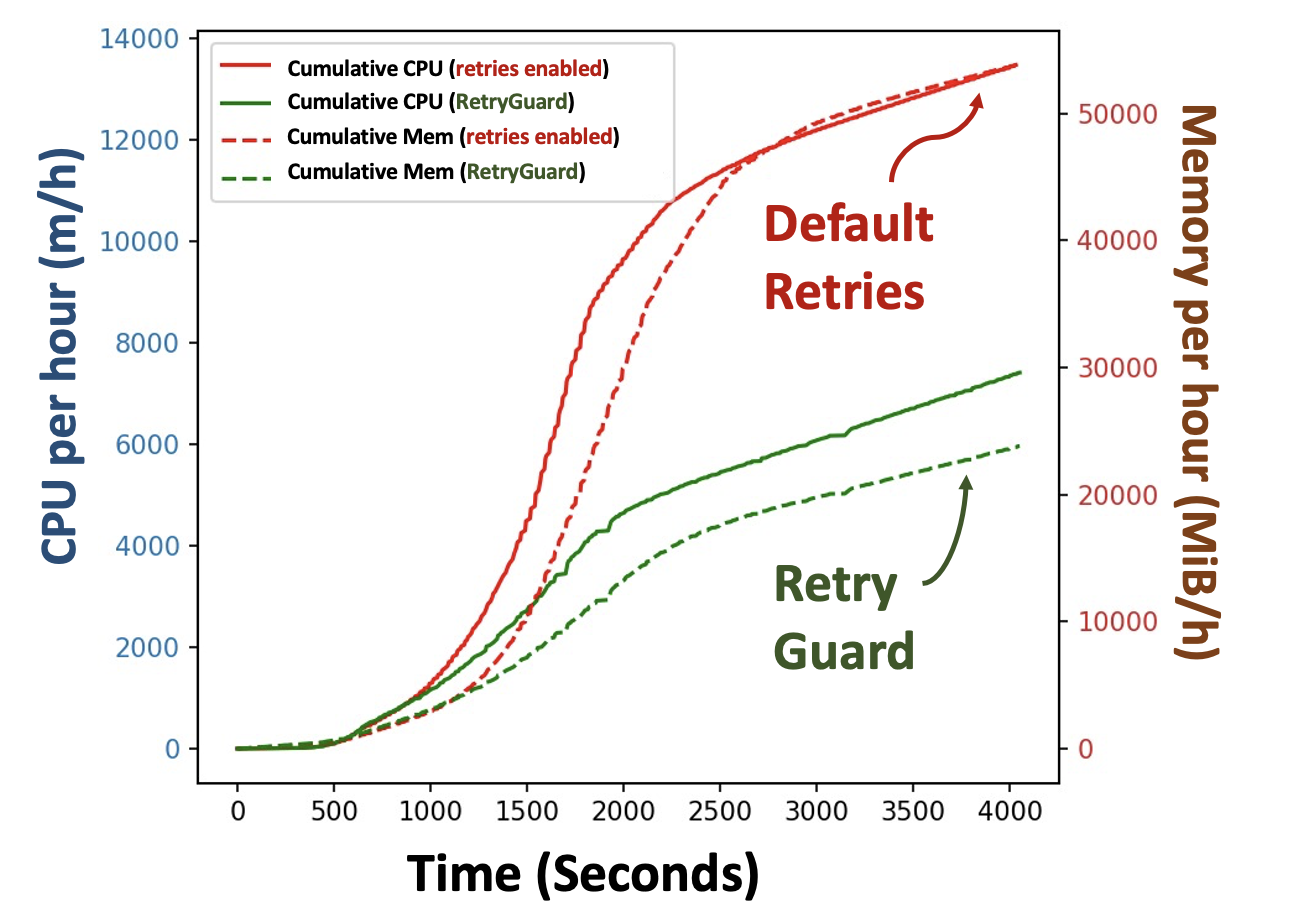}
    \caption{Cumulative CPU and memory.}
    \label{fig:cumulative_solution}
    \end{subfigure}
\caption{Results from the Istio Bookinfo experiments.}
    \label{fig:exp_istio}
\end{figure}

Enabling \emph{RetryGuard} effectively prevents the retry storm that occurs when retries are enabled. With \emph{RetryGuard}, the number of requests received by the \textit{Reviews} service stabilizes, and the service itself scales up more moderately. 
Although only \textit{Reviews} was configured with a slower HPA, \emph{RetryGuard}'s benefits of containing the retry storm propagated to the entire system, including the \textit{Ratings} and \textit{Product} services.
\emph{RetryGuard} further lowers total resource consumption, as reflected in the lower number of replicas, CPU usage, and memory usage.
With \emph{RetryGuard}, the system peaks at 209 replicas, compared with
616 replicas without it (see Fig.~\ref{fig:num_pods_solution}), a nearly
$3\times$ reduction in peak replica count. Over the entire experiment,
\emph{RetryGuard} also reduces cumulative memory usage by 55\%
(see Fig.~\ref{fig:cumulative_solution}).

Performance actually improves: under the default configuration, the rejection rate was 5\%, and the average latency was 5.17 seconds. With \emph{RetryGuard}, the rejection rate was reduced to 4\% and latency dropped to 4.02 seconds. Hence, we reduced both time and cost, without compromising reliability.

We also observed that \textit{RetryGuard} only disabled retries for the \emph{Reviews} service, which frequently encountered errors. 
Meanwhile, the \emph{Product}, \emph{Details}, and \emph{Ratings} services maintained retries almost continuously, preserving the benefits of retry mechanisms in these services.

\section{Discussion}\label{sec:5}
\label{sec:discussion}

\begin{figure}[h]
	\centering
\includegraphics[width=\linewidth]{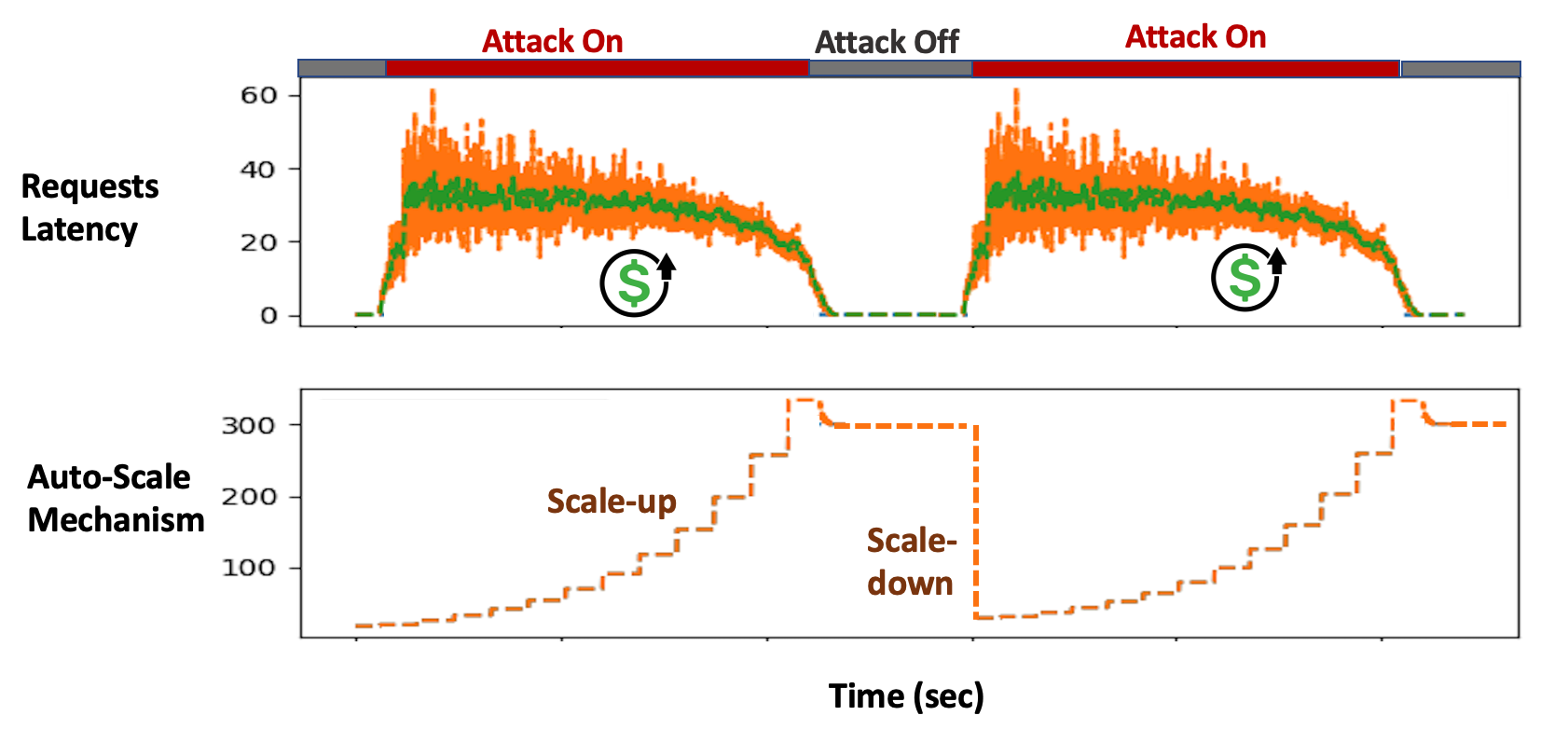}
\caption{Burst-DDoS: Retry storms amplify the attack's impact and may be repeated to exhaust the victim's budget.}
	\label{fig:selfddosexpsecurity}
\end{figure}

\noindent
\textbf{Malicious Attacks: Short DDoS Burst, Long-Term DoW.}\label{sec:5.2}
Applications that employ retry patterns are at significant risk of substantial economic damage from  DDoS attacks. 
Recently, attacks involving short, repeating waves have become increasingly common, confounding conventional DDoS mitigation and detection solutions \cite{alcoz2022aggregate, yoyoinfocom2017, li2023comprehensive}. In many cases, the goal is to strain the budget of the victim.
When retries are enabled, any miscoordination in scaling is likely to trigger a self-inflicted DoW ``snowball'' scenario that persists for minutes after each short-burst attack, even if the attack itself lasts only a few seconds. As the attacker halts the attack burst, malicious requests remain in the system, forming a queue of pending requests. Consequently, new incoming requests are impacted and conflict with the retried malicious requests, creating a rolling effect of queued requests. This cascading issue persists until the system scales up and clears the queue. See Fig. \ref{fig:selfddosexpsecurity}.

\vspace{0em}

Using \textit{RetryGuard}, the system detects overload within one interval and prevents retries from amplifying the attack beyond the attacker’s own traffic.
\\

\noindent
\textbf{Economic Interplay.} 
It is important to note that the overall operational cost of microservices in the cloud is a rather intricate issue. This results from the interplay between the economic players. The cloud clients (namely application owners) find the serverless services approach attractive as they seek to concentrate on the application level and avoid getting into the details of the services. 
However, as we demonstrate, this approach may expose application owners to significant added costs, stemming from the complex interactions among their microservices.
Cloud providers, on the other hand, may have little incentive to address this problem, as the excess usage it generates ultimately adds to users' bills, and thus to their own revenues.

\section{Concluding Remarks}\label{sec:conclusions}
We introduced \textit{RetryGuard}, a distributed framework that
suppresses counterproductive retries during prolonged miscoordination.
Our analysis and evaluation show that \textit{RetryGuard} substantially
reduces resource consumption and cost. In our AWS Serverless deployment,
it achieves a 98\% reduction in storm size and more than a 90\% reduction
in latency; in our Kubernetes Istio deployment, it reduces peak replicas
by $3\times$ and cumulative memory usage by 55\%. Future work can explore
broader forms of regime-aware retry control.

\bibliographystyle{IEEEtran}
\bibliography{Retries_DoW_SIGCOMM_2025}

\end{document}